\documentclass[12pt,preprint]{aastex}

\shorttitle{The AGN nature of 11 out of 12 \emph{Swift}/\emph{RXTE} unidentified sources}
\shortauthors{Landi et al.}
\begin{document}

\title{The AGN nature of 11 out of 12 \emph{Swift}/\emph{RXTE} unidentified sources through optical 
and X-ray spectroscopy}

\author{R.~Landi\altaffilmark{1}, N.~Masetti\altaffilmark{1},
L.~Morelli\altaffilmark{2}, E.~Palazzi\altaffilmark{1},
L.~Bassani\altaffilmark{1}, A.~Malizia\altaffilmark{1},
A.~Bazzano\altaffilmark{3}, A.J.~Bird\altaffilmark{4},
A.J.~Dean\altaffilmark{4}, G.~Galaz\altaffilmark{2},
D.~Minniti\altaffilmark{2}, and P.~Ubertini\altaffilmark{4}}

\altaffiltext{1}{INAF -- Istituto di Astrofisica Spaziale e Fisica Cosmica di Bologna, Via 
P. Gobetti 101, I-40129 Bologna, Italy}
\altaffiltext{2}{Departamento de Astronom\'{\i}a y Astrof\'{\i}sica, 
Pontificia Universidad Cat\'olica de Chile, Casilla 306, Santiago 22, 
Chile}
\altaffiltext{3}{INAF -- Istituto di Astrofisica Spaziale e Fisica Cosmica di
Roma, Via Fosso del Cavaliere 100, I-0033 Roma, Italy}
\altaffiltext{4}{School of Physics and Astronomy, University of 
Southampton, Highfield, Southampton, SO17 1BJ, UK}

\begin{abstract}
The \emph{Swift} Burst Alert Telescope (BAT) is performing a high Galactic latitude survey in 
the 14--195 keV band at a flux limit of $\sim$$10^{-11}$ erg cm$^{-2}$ s$^{-1}$, 
leading to the discovery of new high energy sources, most of which have not so far been properly 
classified. A similar work has also been performed with the \emph{RXTE} slew survey leading to the 
discovery of 68 sources detected above 8 keV, many of which are still unclassified.
Follow-up observations with the \emph{Swift} X-ray Telescope (XRT) provide, for many of these objects,
source localization with a positional 
accuracy of few arcsec, thus allowing the search for optical counterparts to be more efficient
and reliable.
We present the results of optical/X-ray 
follow-up studies of 11 \emph{Swift} BAT detections and one AGN detected in the 
\emph{RXTE} Slew Survey, aimed at identifying their longer-wavelength counterparts and at 
assessing their nature.
These data allowed, for the first time, the optical classification of 8 objects and a
distance determination for 3 of them. For another object, a more refined optical classification
than that available in the literature is also provided.
For the remaining sources, optical spectroscopy provides a characterization of the source
near in time to the X-ray measurement.
The sample consists of 6 Seyfert 2 galaxies, 5 Seyferts of intermediate type 1.2--1.8, 
and one object of Galactic nature - an Intermediate Polar (i.e., magnetic) Cataclysmic 
Variable. Out of the 11 AGNs, 8 ($\sim$70$\%$) including 2 Seyferts of type 1.2 and 1.5, are absorbed 
with $N_{\rm H} > 10^{22}$ cm$^{-2}$.
Up to 3 objects could be Compton thick (i.e. $N_{\rm H} > 1.5 \times 10^{24}$ cm$^{-2}$),
but only in one case (Swift J0609.1--8636) does all the observational evidence strongly suggests this 
possibility. The present data demonstrate the capability of coordinated hard X-ray 
and optical observations to discover absorbed AGNs.
\end{abstract}


\section{Introduction}

Recently, significant progress has been made in surveying the extragalactic 
sky at photon energies above 10 keV. Observations at these energies are efficient 
for finding absorbed AGNs, as they probe heavily obscured regions/objects, i.e. 
those that could be missed in optical, UV and even soft X-ray surveys.

Quantifying the fraction of nearby absorbed AGNs, particularly those that have 
$N_{\rm H}> 1.5\times10^{24}$ cm$^{-2}$ (i.e. the ones in Compton thick regime),
is necessary if one wants 
to understand the accretion history of the Universe and study a population 
of objects poorly explored so far. Furthermore, the distribution of column 
densities is a key parameter for estimating the contribution of AGNs to 
the X-ray cosmic diffuse background and for testing current unified 
theories. A number of surveys performed above $\sim$10 keV are now available and can
be used to study absorption in AGNs.  

The recent \emph{RXTE} survey made in the 8--20 keV energy band using slew 
observations (Revnivtsev et al. 2004; Sazonov \& Revnivtsev 2004)
resulted in a list of 68 objects (above 3$\sigma$ level) down to a flux 
limit of about $3.4\times10^{-11}$ erg cm$^{-2}$ s$^{-1}$; 
14 of these detections are still unidentified and so lack optical 
classification, which is the first step in any survey work (but see Bikmaev et 
al. 2006). Furthermore, no information on the X-ray spectral shape of the less 
known objects is available. 

An even greater step forward has been provided by the 
\emph{Swift} BAT survey, which is sensitive in the 15--195 keV band above 
a flux of $\sim$$10^{-11}$ erg cm$^{-2}$ s$^{-1}$ (Markwardt et al. 2005).
Although primarily devoted to the study of Gamma-Ray Bursts, \emph{Swift} 
observes the sky in survey mode when it is not responding to these events; 
with a large field of view of $>$ 2 sr, galactic and extragalactic 
X-/$\gamma$-ray sources can thus be serendipitously detected.
Preliminary results from the first three months of observations provided a 
first sample of 66 objects: 12 were Galactic sources and 45 extragalactic 
objects (44 AGNs and the Coma cluster); the 9 remaining sources were
not optically classified.
After this initial work, a number of new detections have also been reported 
(Tueller et al. 2005; Kennea et al. 2005; Ajello et al. 2006a) and often 
the proposed counterpart is not optically classified or 
characterized in X-rays below 10 keV.

In parallel with our ongoing program of optical 
classification of \emph{INTEGRAL} IBIS sources (Masetti et al. 2004, 
2006a,b,c,d), we have included in our activities follow-up work on 
unclassified objects detected with \emph{Swift} and \emph{RXTE}.
Here, we report on optical/X-ray follow-up observations of a sample of 
11 \emph{Swift} BAT detections (Tueller et al. 2005; Kennea et al. 
2005; Ajello et al. 2006a,b; Grupe et al. 2006) plus of an AGN detected 
in the \emph{RXTE} Slew Survey (Revnivtsev et al. 2006). 
Our optical spectroscopy allows for the first time the classification of 8 objects and
a distance estimate for 3 of them; for another object a more refined optical classification 
than that available in the literature is also provided. Despite having already been classified in the literature, 3 more objects were re-observed 
in order to have information on their optical status closer in time to the X-ray measurements.
Although for most, but not all, objects in the sample, X-ray spectroscopy was 
published in the form of Atels, we have repeated the analysis here in order to 
include more observations and to provide a more uniform data set.
In some cases we find significant differences between the published results
and our own analyses.
In the case of Swift J0918.5+1618 we also retrieved and analyzed an archival
\emph{ASCA} observation in order to have more detailed information on the source X-ray spectrum 
and to solve the issues related to the possible Compton thick nature of this type 1 Seyfert
(Ajello et al. 2006a).

The optical results presented in this 
paper and concerning the \emph{Swift} sources supersede the preliminary 
and concise ones presented in the Atels by Masetti et al. (2006e) and Morelli et al. (2006).

\section{X-ray observations and data analysis}

In this Section we present X-ray observations acquired with the XRT (X-ray Telescope, 0.2--10
keV) on board the \emph{Swift} satellite (Gehrels et al. 2004). 

\begin{figure}
\centerline{\epsscale{0.7} \plotone{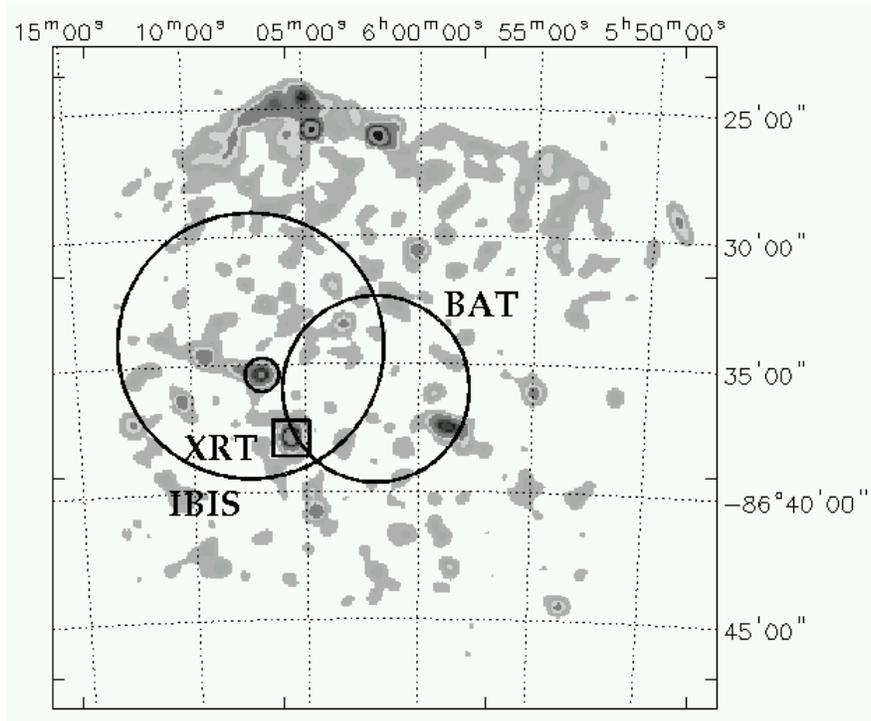}}
\caption{XRT field of view of the region surrounding Swift J0601.9--8636. The two large circles
represent the \emph{INTEGRAL} IBIS (left) and \emph{Swift} BAT (right) error boxes, 
while the XRT position of the source is given by the small box. The smaller circle shows instead the 
position of the other object detected nearby and classified as a normal star.}
\label{errorbox}
\end{figure}

XRT data reduction was performed using the XRTDAS v. 1.8.0 standard data 
pipeline package ({\sc xrtpipeline} v. 0.10.3), in order to produce 
screened event files. All data were extracted only in the Photon Counting 
(PC) mode (Hill et al. 2004), adopting the standard grade filtering (0--12 
for PC) according to the XRT nomenclature. The log of all X-ray
observations is given in Table~\ref{Tab1}. For each measurement, we report the position of
the XRT source associated with the high energy emitting object seen with \emph{Swift} BAT
or with \emph{RXTE} PCA, its uncertainty at a 90\% confidence level, the XRT observation date, the 
relative exposure time,
the energy band adopted in the source spectral fitting (which may vary depending on the
statistical quality of each exposure), and the corresponding source count rate.
Images have been extracted in the 0.3--10 keV energy band and searched for significant excesses 
associated with the \emph{Swift} BAT or \emph{RXTE} PCA sources.
In all cases, a single bright X-ray source was detected inside the BAT positional 
uncertainty; for the \emph{RXTE} source, the bright object detected with XRT coincides with the 
proposed \emph{ROSAT} counterpart (Revnivtsev et al. 2006).
The only exception, Swift J0601.9--8636, was $3^{\prime}.6$ away from the BAT detection (Kennea 
et al. 2006), i.e. just marginally consistent with the BAT position. 
This source has recently been detected also by \emph{INTEGRAL} IBIS
and localized with a $5^{\prime}.2$ uncertainty by this instrument (Bird et al. 2007);
the intersection of \emph{Swift} BAT and \emph{INTEGRAL} IBIS error boxes clearly 
indicates that only LEDA 18394 (also ESO 005--G004, box in Figure~\ref{errorbox}) is the counterpart 
of this high energy emitting object.
The other possible counterpart (smaller circle in Figure~\ref{errorbox})
can be excluded as optical spectroscopy
acquired with the same setup at the 1.5m CTIO telescope on February 19, 2007, shows that it is a
star, likely of spectral type A, without any peculiar spectral characteristics, i.e. it is
unlikely to emit X-rays above 10 keV.

Events for spectral and temporal analysis 
were extracted within a circular region of radius 20$^{\prime \prime}$, 
centered on the source position,
which encloses about 90\% of the PSF at 1.5 keV (see Moretti et al. 2004).
The background was taken from various source-free regions close to the 
X-ray source of interest, using both circular/annular regions with 
different radii in order to ensure an evenly sampled background. In all 
cases, the spectra were extracted from the corresponding event files using 
the {\sc XSELECT} software and binned using {\sc grppha} in an appropriate 
way, so that the $\chi^{2}$ statistic could be applied. We used the 
latest version (v.008) of the response matrices and create individual 
ancillary response files (ARF) using {\sc xrtmkarf}. Spectral analyses 
have been performed using XSPEC version 12.2.1, while for temporal analyses
we adopted the XRONOS package (version 5.21).

Each individual observation was inspected in order to assess the presence
of short-term variability. In only three cases did we find 
significant short-term changes (Swift J0444.1+2813, Swift J0732.5--1331 and 
Swift J1200.8+0650 with a rms variability of 20--40\%, 10--30\% and $\sim$20\%, respectively), 
but the low statistical quality of the data prevented any spectral analysis 
within segments of these observations.

For sources with more than one pointing \footnote{In a few cases some observations have not been
considered because their poor statistics did not allow us to perform a reliable spectral
analysis.}, we first analysed individual spectra to search for variability, and then 
combined the data to improve the statistics.
 
In just a few cases (Swift J0918.5+1618, Swift J1009.3--4250 and
Swift J1038.8--4942), we found significant changes in flux, but not in spectral shape, 
between different observing periods.
To allow for these flux variations, fits to the combined data
were performed with the power law normalization free to vary.

Due to the limited spectral signal often available, 
we first employed a simple model of a power law absorbed by both a Galactic (Dickey 
\& Lockman 1990) and an intrinsic column density (our baseline). 
If this baseline model was not sufficient to fit the data, we then introduced extra spectral 
components as required.

The results of the analysis of both single and combined observations are 
reported in Table~\ref{Tab2} and Table~\ref{Tab3} respectively, where we list: 
the Galactic absorption according to Dickey \& Lockman 
(1990); the column density in excess to the Galactic value; the photon 
index; the 2--10 keV flux; the reduced $\chi^{2}$ of the best-fit model and 
the column density measured by fixing the spectral photon index to 
a canonical value for an AGN ($\Gamma=1.8$). We also list in Table~\ref{Tab3} 
the parameters relevant to any extra spectral features (e.g. soft excess component and iron line) 
required by the data in a few sources. In particular, we have added: an unabsorbed 
power law component having the same photon index of the
primary continuum in 3 type 2 AGNs (Swift J0444.1+2813, Swift J1009.3--4250 and Swift J1930.5+3414); 
a thermal component in the only galactic source in the sample (Swift J0732.5--1331); and an iron 
line in two AGN 
(Swift J1238.9--2720 and XSS J12303--4232). The extra power law could be interpreted as
scattered emission, which is a spectral component often found in type 2 Seyferts (Risaliti 2002).
All quoted errors correspond to a 90$\%$ confidence level for a single
parameter of interest ($\Delta\chi^{2}=2.71$).

In one case (Swift J0918.5+1618), we also used archival 
\emph{ASCA} GIS (Gas Imaging Spectrometers, Ohashi et al. 1996) data to better 
characterize the source spectrum. The relevant spectra and associated files were downloaded from 
the TARTARUS database\footnote{available at\\{\tt http://tartarus.gsfc.nasa.gov/}} and analyzed with 
the same XSPEC version used for the XRT data. For this particular case, the best-fit results are
presented in Table~\ref{Tab4}. 

Figures 2 to 12 show, for those AGN in the sample with high signal-to-noise ratio,
the results of this spectral analysis, which are discussed in more details in the subsections devoted to 
each individual object. The X-ray spectra of the only galactic source in the sample 
(Swift J0732.5--1331) are presented and discussed in the Appendix.

\section{Optical observations and data analysis}

Sources with Dec $>$ $-$20$^{\circ}$ were observed spectroscopically 
at the Bologna Astronomical Observatory in Loiano (Italy) with the 1.5-metre 
telescope ``G.D. Cassini'' equipped with the BFOSC instrument, which uses a 
1300$\times$1340 pixel EEV CCD. In all observations, Grism \#4 and a slit 
width of $2^{\prime \prime}$ were used, providing a 3500--8700 \AA~nominal spectral 
coverage. The use of this setup secured a final dispersion of 4.0~\AA/pix 
for all these spectra. Spectroscopic observations of the remaining 
southern sources were secured with the 1.5-metre CTIO telescope of Cerro 
Tololo (Chile) equipped with the R-C spectrograph, which carries a 
1274$\times$280 pixel Loral CCD. Data were acquired using Grating \#13/I 
and with a slit width of 1$\farcs$5, giving a nominal spectral coverage 
between 3300 and 10500 \AA~and a dispersion of 5.7~\AA/pix. 
In one case (Swift J1009.3--4250 = ESO 263--13) we retrieved an archival 
optical spectrum, with dispersion 1.6~\AA/pix, acquired with the 4-metre 
Anglo-Australian Telescope (AAT) within the 6dF Galaxy Survey (Jones 
et al. 2004); we refer the reader to this latter paper for observational 
details. The complete log of the observations is reported in 
Table~\ref{Tab5}. 

After cosmic-ray rejection, the Loiano and CTIO spectra were reduced, background
subtracted and optimally extracted (Horne 1986) using IRAF\footnote{IRAF
is the Image Reduction and Analysis Facility made available to the
astronomical community by the National Optical Astronomy Observatories,
which are operated by AURA, Inc., under contract with the U.S. National
Science Foundation. It is available at {\tt http://iraf.noao.edu/}}.
Wavelength calibration was performed using He-Ar lamps acquired soon
after each spectroscopic exposure; the Loiano spectra were then 
flux-calibrated using the spectrophotometric standard BD+25$^\circ$3941 
(Stone 1977), whereas the CTIO spectra used the standard star LTT 7379
(Hamuy et al. 1992, 1994).
Finally, and when applicable, different spectra of the same object were
stacked together to increase the S/N ratio. The wavelength calibration
uncertainty was $\sim$0.5~\AA~for all cases; this was checked using
the positions of background night sky lines.
Instrumental broadening of spectral features is (in terms of $\sigma$
assuming a Gaussian description) $\approx$260 km s$^{-1}$ and 
$\approx$300 km s$^{-1}$ for the Loiano and CTIO spectra, respectively.
 
To flux calibrate the optical spectrum of Swift J1009.3--4250 (originally expressed
in counts) we considered the $B$ and $R$ magnitudes of Winkler \& Payne 
(1990), which were acquired on a 30$''$ circular region centered on the 
source, and rescaled them to the 6dF optical fibre aperture (6$\farcs$7 
in diameter). This was done using the SuperCosmos (Hambly et al. 2001)
digitized images and the conversions from the SuperCosmos to the 
Johnson-Cousins photometry system (Colless 2003).
We then used these rescaled magnitudes to convert the spectrum
counts into physical units. For the AGN classification, 
we used the criteria of Veilleux \& Osterbrock 
(1987) and the line ratio diagnostics of Ho et al. (1993, 1997); moreover, 
for the subclass assignation of Seyfert 1 nuclei, we used the 
H$_\beta$/[O {\sc iii}]$\lambda$5007 line flux ratio criterion as in 
Winkler (1992).

The quoted optical magnitudes are extracted from the USNO-A2.0\footnote{available at\\ 
{\tt http://archive.eso.org/skycat/servers/usnoa/}} catalog.
Table~\ref{Tab6} reports the emission line fluxes of the most interesting
lines in each spectrum.
The line fluxes and absolute magnitudes from extragalactic sources
presented in this paper were dereddened for Galactic absorption along
their line of sight following the prescription for the computation of the
Galactic color excess $E(B-V)_{\rm Gal}$ (also reported in Table~\ref{Tab6})
as in Schlegel et al. (1998) and assuming the Galactic extinction law of
Cardelli et al. (1989). For the intrinsic reddening calculations, we
assumed an intrinsic H$_\alpha$/H$_\beta$ line ratio of 2.86 (Osterbrock
1989). The spectra of the galaxies shown here were not corrected for
starlight contamination (see, e.g., Ho et al. 1993, 1997) given their
limited S/N and resolution. We do not consider this to affect any of our
conclusions. We have also compared, in Table~\ref{Tab6}, the extinction obtained from the optical 
($A_{\rm V_{\rm opt}}$) and X-ray data ($A_{\rm V_{\rm X}}$).
When this is possible, we find that $A_{\rm V_{\rm opt}}$ is generally smaller than $A_{\rm V_{\rm X}}$;
for AGN this is not a new result as it was first
noticed by Maccacaro et al. (1982). The cause of this discrepancy is not clear, but 
could be due to the optical and X-ray absorption coming from different media/regions 
(Weingartner $\&$ Murray 2002)
or due to a dust to gas ratio or dust grain dimensions different than in our galaxy (Maiolino et al. 
2001). Only in one case (Swift J0601.9--8636) do we find that $A_{\rm V_{\rm X}}$ is smaller than
$A_{\rm V_{\rm opt}}$, but this may be due a false estimate of the absorption column density
in the X-ray spectrum (see Section 5.2) and the likely Compton thick nature of this source
(see Section 5.2).

Figures 13 and 14 show the optical spectra of all AGN in the sample, while the optical spectrum
of the galactic source Swift J0732.5--1331 is presented in the Appendix.

\section{Diagnostic ratios}

To further characterize our objects, we have used diagnostic diagrams to confirm 
the presence of an active nucleus and to discriminate between Compton thin and thick sources.
It is in fact possible in some cases that the AGN component is so deeply absorbed that other
components, such as a starburst, dominate in the optical spectrum (Masetti et al. 2006c).
Since the IR emission is generally associated with 
star-forming activity while the [O {\sc iii}] emission is mainly produced by 
photons generated in the active nucleus, we can employ the 
$L_{\rm [OIII]}$/$L_{\rm IR}$ ratio to discriminate between AGN and Starburst galaxies. 
Indeed, in the Ho et al. (1997) sample of nearby galaxies, 90$\%$ of those classified as starbursts  
have a value of $L_{\rm [OIII]}$/$L_{\rm IR}$ below 
10$^{-4}$, while 88$\%$ of the AGN have ratios above this value (Panessa and Bassani 2002).

Generally, indirect arguments are used to probe the Compton thick versus thin nature of 
a type 2 AGN, and these include the equivalent width of the iron line, and the ratio of anisotropic 
(i.e. modified by absorption such as the 2--10 keV emission) versus isotropic (i.e. not modified by 
absorption) luminosities.
The [O {\sc iii}]$\lambda$5007 ([O {\sc iii}] hereafter) flux is 
considered as a good isotropic indicator because it is produced in the 
Narrow Line Regions, which are not strongly affected by absorption (Maiolino \& Rieke 1995; Risaliti et al. 
1999; Bassani et al. 1999). The $L_{\rm X}$/$L_{\rm [OIII]}$ ratio has been studied in a 
large sample of Seyfert 2 galaxies; all Compton thin Seyfert (type 1 and 
2) show ratios higher than 1, while Compton thick sources show $L_{\rm 
X}$/$L_{\rm [OIII]}$ ratios below this value (Bassani et al. 1999).

Another isotropic indicator is the Far-Infrared (FIR) emission which is probably produced 
in even more external regions; typically, type 1 and Compton thin type 2 AGNs show 
$L_{\rm X}$/$L_{\rm IR}$ ratios around or above 0.1, while Compton thick type 2 objects show 
ratios lower than $5\times10^{-4}$ (David et al. 1992; Mulchaey et al. 1994; Risaliti et al. 1999).
However, the validity of this diagnostic is less certain because many components 
besides the nuclear one
(for example coming from star-burst activity and from the host galaxy) may contribute to the IR flux.
These caveats should be kept in mind when dealing with this particular ratio.

Finally, emission above 20 keV can be used as an isotropic probe too as it is virtually unaffected
by absorption as long as logN$_{\rm H}$ is below 24.5; for higher column density even the emission
above 20 keV is blocked and can only be seen indirectly.
Using a sample of hard X-ray selected AGN, Malizia et al. (2007) found that the
$L_{\rm X}$/$L_{\rm HX}$ ratio of type 1 AGN is typically around 0.7, decreasing 
progressively 
as the absorption increases; Compton thick objects are easily recognized for having 
negligible absorption and a ratio below 0.02.

In conclusion, these four ratios can provide an independent way to 
establish which is the dominant component between AGN and Starburst and, at 
the same time, they are a powerful tool in the detection of Compton thick 
sources when the X-ray spectral analysis is not sufficient to recognize them. 
Here, the optical spectra provide information on the [O {\sc iii}] 
flux and the X-ray spectroscopy on the X-ray one. 
Hard X-ray fluxes in the 10--100 keV energy band are instead obtained from available
\emph{Swift} BAT observations (see references in Table~\ref{Tab7}).
When possible, the [O {\sc iii}] flux of each galaxy has been corrected for extinction assuming 
the absorption local to the narrow-line region as determined from the 
H$_\alpha$/H$_\beta$ line ratio\footnote{$F_{\rm [OIII],corr} = F_{\rm [OIII],obs}
\left [\frac{ \left (\frac{H_\alpha}{H_\beta} \right )} 
{\left (\frac{H_\alpha}{H_\beta} \right )_{\rm 
0}} \right ]^{2.94}$, where $\left (\frac{H_\alpha}{H_\beta} \right )_{\rm
0}=3$} (Bassani et al. 1999); the X-ray and hard X-ray fluxes are instead not corrected for absorption. 
IR fluxes have been estimated from \emph{IRAS} data extracted 
from the NASA/IPAC Extragalactic Database\footnote{available at
{\tt http://nedwww.ipac.caltech.edu/}} (NED) archive or from the HEASARC archive.
For this paper, we adopt the same definition used in Mulchaey et al. 
(1994) to estimate the IR flux: $F(\rm IR)=S_{25~\mu 
m}\times(\nu_{25~\mu m})+S_{60~\mu m}\times(\nu_{60~\mu m})$. 
The IR fluxes together with the hard X-ray fluxes estimated in this way, as well as the luminosity 
ratios used in the present study are all listed in Table~\ref{Tab7}. 

Using these ratios, we can conclude that most AGN in the sample are dominated by an 
active nucleus rather than by starburst activity. Furthermore, they are mostly characterized 
by being in the Compton thin regime. Three objects could be Compton thick, but only in one case 
(Swift J0601.9+8336) does all the observational evidence strongly suggests this possibility.

\section{Results on individual sources}

{bf In the following subsections, we present X-ray and optical results of all AGN 
analyzed in this paper. The only galactic source found is discussed separately in the Appendix.}
We have assumed a cosmology with $H_{\rm 0}$ = 65 
km s$^{-1}$ Mpc$^{-1}$, $\Omega_{\Lambda}$ = 0.7 and $\Omega_{\rm m}$ = 0.3.

\subsection{Swift J0444.1+2813}

\begin{figure}
\rotatebox{-90}{\epsscale{0.7} \plotone{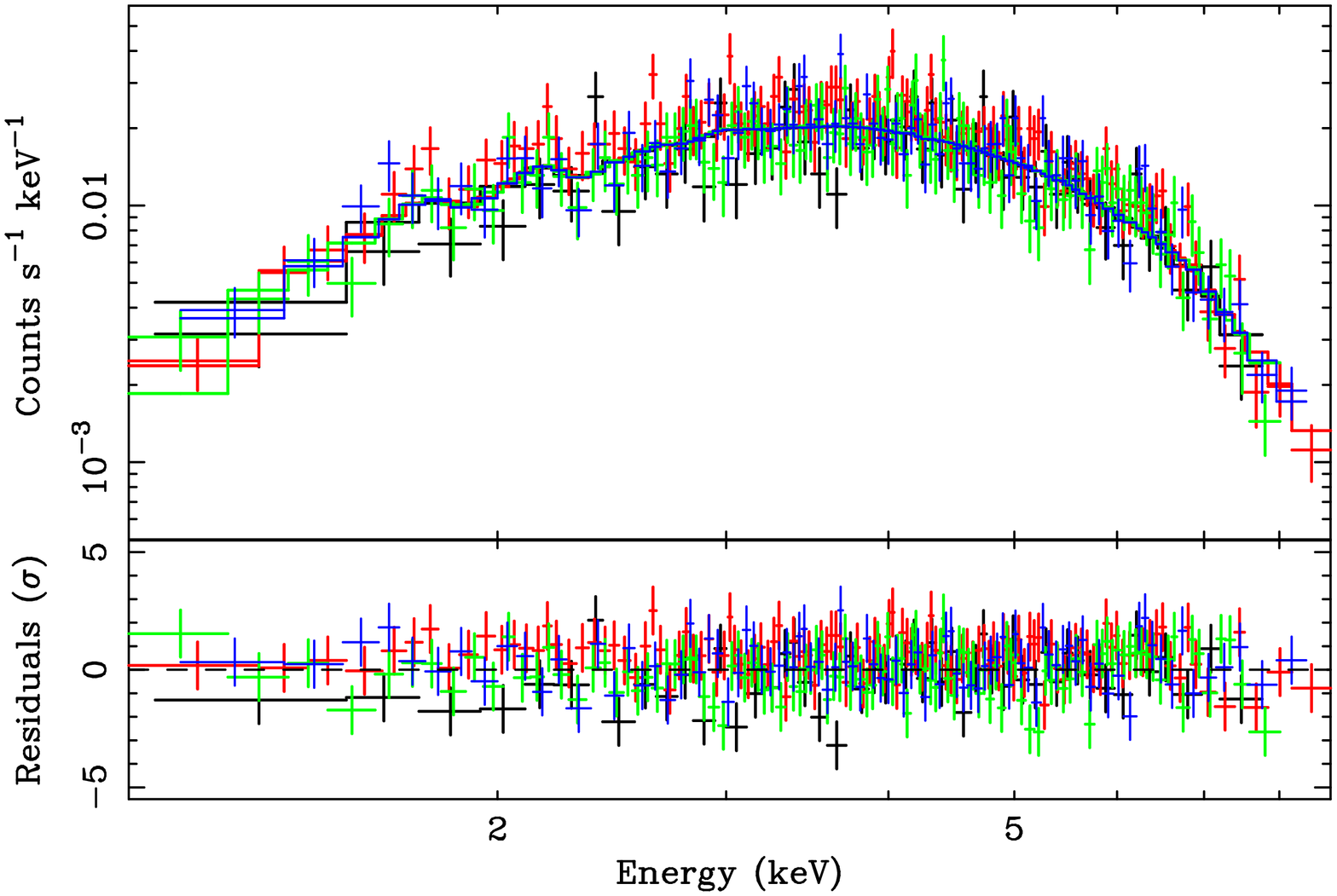}}
\caption{Four XRT spectra of Swift J0444.1+2813 fitted with a primary absorbed power
law component plus a power law, having the same photon index, and absorbed only by Galactic
column density (upper panel); residuals to this model in units of $\sigma$ (lower panel).}
\label{spec0444}
\end{figure}

This BAT source is identified with the 2MASS (Two Micron All Sky Survey, 
Skrutskie et al. 2006) extended object 2MASX J04440903+2813003 (Tueller et 
al. 2005), associated with an unclassified principal galaxy (PGC 86269 at $z=0.011268$; 
Pantoja et al. 1997). It is listed in the NVSS (National Radio Astronomy 
Observatory [NRAO] Very Large Array [VLA] Sky Survey Catalog; Condon et 
al. 1998) with a 20 cm flux of $28.6\pm0.9$ mJy. The object is also an 
\emph{IRAS} source 
and has never been detected at X-ray energies before. A preliminary 
analysis of the XRT data indicated heavy absorption ($N_{\rm H}$ $\sim$$10^{23}$ cm$^{-2}$),
but did not provide any information on the source spectral shape (Tueller et al. 2005).
The XRT spectra are all 
well fitted with two power laws having the same photon index but with only one intrinsically absorbed 
(see Table~\ref{Tab3} and Figure~\ref{spec0444}).
The addition of an unabsorbed
power law component takes into account excess emission below 2 keV and is required 
by the data with a significance greater than the 99.73$\%$ confidence level in each observation
($>$ 99.99$\%$ for the combined spectrum).
The primary power law is flat and absorbed by an intrinsic column density of 
$\sim$$4\times10^{22}$ cm$^{-2}$. The absorption is lower 
than the preliminary value reported by Tueller et al. (2005).

Optical spectroscopy of the object shows the presence of narrow 
H$_\alpha$, [N {\sc ii}], [S {\sc ii}], [O {\sc i}] and [O {\sc iii}] 
emission lines, with Full Width at Half Maximum (FWHM) $\la$ 800 km 
s$^{-1}$ (Figure~\ref{optical}, upper left panel), superimposed on a reddened continuum. 
All features are at redshift $z=0.011\pm0.001$, consistent with the 21 
cm radio measurement of Pantoja et al. (1997). This corresponds to a 
luminosity distance $D_L$ = 51.2 Mpc and gives a luminosity of
(3.1--3.8)$\times10^{42}$ erg s$^{-1}$ in the 2--10 keV energy band. 
The narrowness of the lines and their relative strengths indicate 
that this is a Seyfert 2 galaxy. The 3$\sigma$ upper limit on the Balmer decrement ($>$ 74)
implies a dereddened [O {\sc iii}] flux 
$F_{\rm [OIII]} > 1.4\times10^{-9}$ erg cm$^{-2}$ s$^{-1}$. 
The latter measurement, if compared with the X-ray flux, implies that this source is 
in the Compton thick regime (see Table~\ref{Tab7}); however, the $L_{\rm X}/L_{\rm IR}$
and $L_{\rm X}/L_{\rm HX}$
ratios are typical of Compton thin sources and the upper limit ($EW < 110$ eV)
on the 6.4 keV iron line,
which can be obtained from the X-ray data, excludes a Compton thick object;
also the short-term variability seen in the XRT data argues against a Compton thick nature.

\subsection{Swift J0601.9--8636}

Despite being a principal galaxy, the optical counterpart of this \emph{Swift} AGN (LEDA 
18394 also ESO 005--G004), is still unclassified in the optical. 
The source is a radio emitter, being reported in the SUMSS (Sidney 
University Molonglo Sky Survey; Mauch et al. 2003) catalog with a 36 cm 
flux of $119.3\pm4.5$ mJy and as a Parkes source with a 4850 MHz flux of 
$30\pm5$ mJy; it is fairly bright and extended in the infrared, being 
listed both in the \emph{IRAS} bright galaxy sample (Sanders et al. 2003)
and in the extended 2MASS catalog. Up to now, no X-ray data have been reported from this object. 

If we fit the XRT spectrum  with an absorbed power 
law with $\Gamma=1.8$, we do not find any evidence for 
extra absorption in excess of the Galactic value (see Table~\ref{Tab2}). 

From the nucleus of this galaxy, we detect narrow (FWHM $\la$ 650 km 
s$^{-1}$) emission lines of H$_\alpha$, [N {\sc ii}] and [S {\sc ii}] 
superimposed on a continuum resembling that typical of a spiral galaxy 
(Figure~\ref{optical}, upper right panel).
The H$_\alpha$/[N {\sc ii}] flux ratio is consistent with that typical of Seyfert 2 galaxies,
although this observable alone cannot be used to discriminate the nature of this source.
However, the radio emission and the detection above 20 keV
strongly suggest that we are dealing with an active galaxy. 
We are therefore led to conclude that Swift J0601.9--8636 is probably a hidden Seyfert 2 galaxy.
The estimated redshift is $z=0.006\pm0.001$, 
consistent with that measured by Fisher et al. (1995). This redshift corresponds to  
a luminosity distance $D_L$ = 29.7 Mpc, implying $L_{\rm (2-10~keV)} = 
1.2\times10^{40}$ erg s$^{-1}$. This luminosity is unusually low compared to that above 20 keV
($3.9\times10^{42}$ erg s$^{-1}$), suggesting again extreme absorption. 

The non-detection of H$_\beta$ implies a H$_\alpha$/H$_\beta$
flux ratio 3$\sigma$ upper limit of 38.3, which in turn means $A_{\rm V} >$ 8.1 and a 
hydrogen column $N_{\rm H} > 1.5\times10^{22}$ cm$^{-2}$ local to the 
galaxy nucleus, which is somewhat at odds with the result of the 
(admittedly with low S/N) X-ray data, unless the source is in the Compton
thick regime.

The non-detection of [O {\sc iii}], together with the lower limit on the 
reddening local to the source makes the [O {\sc iii}] flux value (or a 
limit thereof) unconstrained, and so are the $L_{\rm IR}$/$L_{\rm [OIII]}$ 
and $L_{\rm X}$/$L_{\rm [OIII]}$ ratios. 
The observed $L_{\rm X}$/$L_{\rm IR}$ and $L_{\rm X}$/$L_{\rm HX}$ ratios are instead well 
constrained and their very small values are compatible with Swift J0601.9--8636
being a Compton thick AGN; although the small $L_{\rm X}$/$L_{\rm IR}$
value is also consistent with a starburst-powered galaxy, the optical spectrum provides a
H$_\alpha$/[N {\sc ii}] ratio compatible with those of type 2 Seyfert galaxy.
Indeed, a recent \emph{Suzaku} observation of this source has confirmed its Compton thick 
nature (Ueda et al. 2007).

\subsection{Swift J0823.4--0457}  

The XRT position of this BAT source (Ajello et al. 2006a) falls exactly 
on the galaxy Fairall 272 (also PGC 023515 and MCG--01--22--006), which
belongs to a peculiar interacting system and is part of a 
medium-poor cluster of galaxies in the constellation Hydra (Fairall 
1979). This galaxy (still optically 
unclassified) shows strong [O {\sc iii}] $\lambda\lambda$4959, 5007 as 
well as narrow H$_\beta$ emission lines, produced in a region wider 
than the galaxy core (maybe chaotic filaments). These 
emission lines, combined with the spectral features observed in the two other 
neighboring interacting galaxies (Fairall 271 and Fairall 273), 
suggest that Fairall 272 hosts an active nucleus of unknown class (the filaments could be 
associated with ejective or explosive processes).

Fairall 272 is also a radio emitter, detected in the NVSS catalog with a 20 cm 
flux of $37.2\pm1.5$ mJy; as none of the other nearby galaxies is
detected at this frequency, it is likely that Fairall 272 is the most 
active galaxy of the group. An \emph{IRAS} source (IRAS 08205--0446), 
located only 0$^{\prime}$.6 away, is likely associated with Fairall 272.

A preliminary analysis of the XRT data indicate that the X-ray spectrum is either flat
with negligible absorption, or a more canonical one ($\Gamma = 2$), but absorbed 
($N_{\rm H}$ $\sim$$10^{23}$ cm$^{-2}$; Ajello et al. 2006a).
The results of our spectral analysis confirm these previous findings: 
a flat ($\Gamma$ $\sim$0.7) unabsorbed power law is a good description of the XRT data 
($\Delta\chi^{2}/\nu=7.8/13$, but since we cannot put any constraint on the photon index, 
we prefer to fix $\Gamma = 1.8$ in the fitting procedure and estimate the absorption local
to the source (see Table~\ref{Tab2}).

The optical spectrum of Fairall 272 shows narrow (FWHM $\la$ 750 km 
s$^{-1}$) H$_\alpha$, H$_\beta$, [N {\sc ii}], [S {\sc ii}], [O{\sc i}] 
and [O {\sc iii}] emission lines over a flat continuum (Figure~\ref{optical}, 
central left panel). These features put the source at redshift $z=0.023\pm0.001$, 
consistent with the measurement of Fouque et al. (1992), which 
implies a luminosity distance $D_L$ = 108 Mpc and a 2--10 keV luminosity 
of $5.9\times10^{42}$ erg s$^{-1}$. In this case also, the 
line shapes and ratios indicate that this is a Seyfert 2 galaxy.
In view of the optical classification an absorbed power law is  preferred to an 
unabsorbed flat spectrum.

The H$_\alpha$/H$_\beta$ flux ratio of 6.1 provides a dereddened [O {\sc iii}] flux 
of $2.4\times10^{-13}$ erg cm$^{-2}$ s$^{-1}$ and the diagnostic ratios
indicate that 
this is a Compton thin AGN (see Table~\ref{Tab7}). This substantially confirms the 
X-ray characterization of the source as a not heavily absorbed AGN.

\subsection{Swift J0918.5+1618}

This \emph{Swift} object coincides with the galaxy Mrk 704 ($z$ = 0.029; 
e.g., Hewitt \& Burbidge 1991), which has been detected earlier in 
X-rays by \emph{ROSAT} (Schwope et al. 2000) and \emph{ASCA}, and is listed 
in the \emph{RXTE} Slew Survey (Revnivtsev et al. 2004). This galaxy is 
detected in the radio by the NVSS survey with a 20 cm flux of $6.1\pm0.5$ mJy; 
it is also an IRAS source and a 2MASS extended object.

The source has been classified in the literature as a Seyfert galaxy but 
of different types ranging from 1 to 1.5 (see for example a compendium in 
Nagao, Taniguchi \& Murayama 2000); it is also considered a peculiar type 
of Seyfert 1 AGN (i.e. a ``polar scattered'' object) as it has 
spectropolarimetric properties consistent with those found in type 2 
Seyfert galaxies despite its type 1 optical classification
(Nagao, Taniguchi \& Murayama 2000). According to recent 
modeling, ``polar scattered'' Seyfert 1 are those viewed at an inclination
very close to 45$^{\circ}$, such that the line of sight passes through the 
outer layers of the torus whereas type 1 and 2 Seyfert galaxies have 
inclinations lower and greater than 45$^{\circ}$ respectively (Smith et 
al. 2004; Hoffman et al. 2005).

This latter fact seems to be more consistent with an intermediate  
optical classification: given that the Broad Line Region 
(BLR) is obscured by a dusty torus, the emission lines would appear with a 
composite profile consisting of both narrow and broad emissions.

The X-ray information on this object also suggests that we are dealing
with a peculiar object. Ajello et al. (2006a) reported inconsistencies
between the BAT and XRT spectra and further suggested, through the
analysis of archival \emph{ASCA} data, that the source might be in the
Compton thick regime, in contrast with its optical classification as
a type 1 AGN.

This prompted our optical follow-up observations in order to assess the 
spectral characteristics of this object relatively close in time to the 
\emph{Swift} measurement.

Due to the low statistical quality of the first XRT observation, only the last three
measurements have been taken into account for the spectral analysis.
All data sets are compatible with each other being well fitted with a
steep ($\Gamma$ $\sim$1.8--1.9) power law having mild absorption (see Table~\ref{Tab2}).
There was a change in flux ($\sim$60\%) between the three XRT observations separated
by a few months, but no variation within the individual measurements.

Broad He {\sc i}, He {\sc ii}, Fe {\sc i}, and Balmer emissions at 
least up to H$_\zeta$ (FWHM$_{\sc H_\beta}$ $\sim$5500 km s$^{-1}$), as 
well as narrow H$_\alpha$, H$_\beta$, [N {\sc ii}], [S {\sc ii}], [O {\sc 
i}], [O {\sc iii}] and [Ne {\sc iii}] emission lines are detected in the 
optical spectrum of Swift J0918.5+1618
(Figure~\ref{optical}, central right panel), which 
moreover shows a blue continuum. The spectrum, and a 
H$_\beta$/[O {\sc iii}] flux ratio of ($\sim$2.3), together imply a Seyfert 1.2 AGN 
classification for this galaxy. From the observed wavelengths of 
narrow features, we determine a redshift $z=0.028\pm0.001$, in agreement 
with previous measurements. This gives a luminosity distance $D_L$ = 131.9 
Mpc to the source.
At a redshift $z$ = 0.028, the 2--10 keV luminosity of Swift J0918.5+1618 was
$1.6\times10^{43}$ erg s$^{-1}$ at the time of the XRT observation.

Clearly, our optical and X-ray results are at odds with the conclusions of Ajello et al. (2006a),
suggesting the use of \emph{ASCA} data to get a more detailed picture of the source X-ray 
spectrum.

\emph{ASCA} observed Swift J0918.5+1618 on May 12 1998 for an effective exposure time,
for the GIS instrument, of $\sim$38 ks.
A fit to the GIS spectra with a simple power law seen through Galactic 
absorption provides a poor fit to the data ($\chi^{2}/{\rm d.o.f.}=247.7/194$) 
and a flat photon index ($\Gamma$ $\sim$1), 
suggesting the presence of more complex absorption.
We therefore applied a partial covering model, as observed in 
other intermediate Seyfert galaxies like Mrk 6 (Malizia et al. 2003), Mrk 
1152 (Quadrelli et al. 2003) and 4U 1344--60 (Piconcelli et al. 2006).

\begin{figure}
\rotatebox{-90}{\epsscale{0.7} \plotone{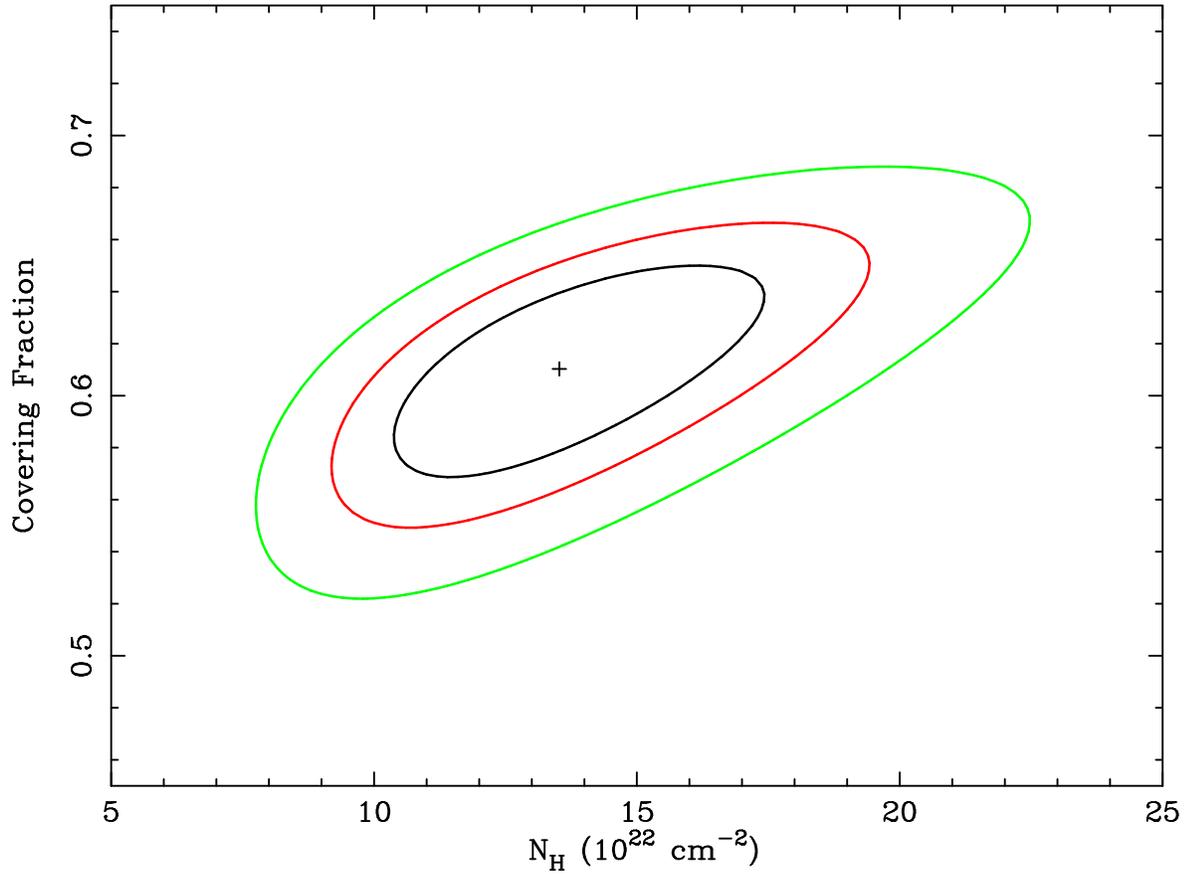}}
\caption{Confidence contours (at 68\%, 90\%, and 99\% confidence level) of the covering
fraction versus column density for the best-fit model used in the \emph{ASCA}
observation of Swift J0918.5+1618.}
\label{spec0918contnh}
\end{figure}

\begin{figure}
\rotatebox{-90}{\epsscale{0.7} \plotone{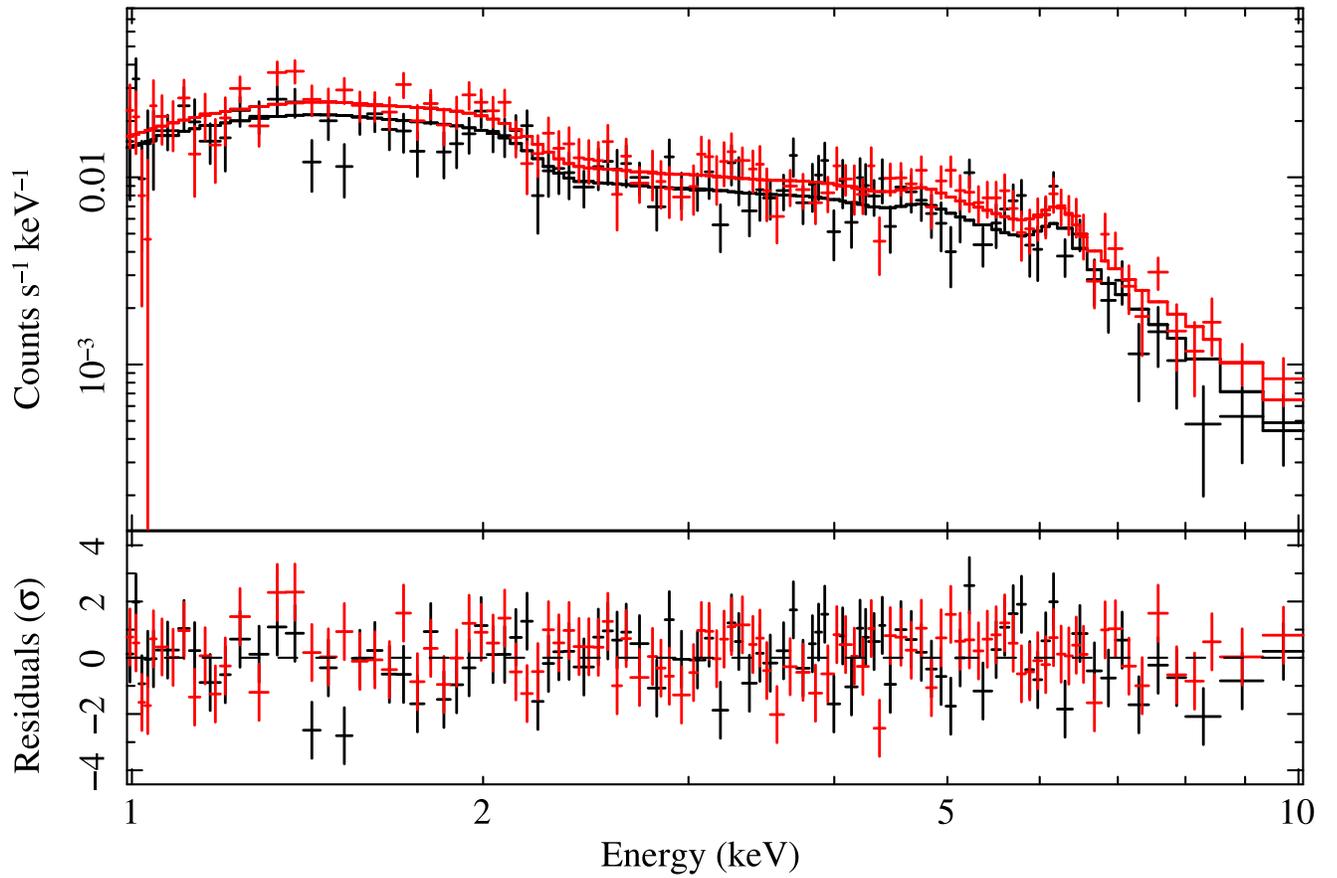}}
\caption{\emph{ASCA} GIS spectra of Swift J0918.5+1618 fitted with a power law
absorbed by a partially covering absorber plus a Gaussian narrow Fe K$\alpha$ line
(upper panel); residuals to this model in units of $\sigma$ (lower panel).}
\label{spec0918asca}
\end{figure}

The addition of this component strongly improves
the $\chi^2$ value, being significant at more than 99.99\% confidence 
level ($\Delta\chi^{2}=43.2$ for two degrees of freedom).
Inspection of the 
residuals around the expected iron line position suggests the 
presence of such a feature. In fact, the inclusion of this extra component
further improves the fit ($\Delta\chi^{2}=11.0$ for two degrees of freedom) and gives a column
density of $\sim$$13\times10^{22}$ cm$^{-2}$ covering $\sim$61$\%$ of the
source (see in Figure~\ref{spec0918contnh} and a narrow line at 6.4 keV with an $EW$ 
of $\sim$233 eV, which is typical of Seyfert 1 galaxies (Turner et al. 1997). 
With this best-fit model (see Table~\ref{Tab4} and Figure~\ref{spec0918asca}), 
the power law photon index is 1.5, similar to that observed in other AGN 
showing complex absorption.
Extrapolation of the \emph{ASCA} best-fit to the 10--100 keV energy band provides
a flux consistent with the BAT one.

In order to search for consistency between the \emph{ASCA} and XRT results,
we re-fitted the whole XRT data set with the \emph{ASCA} best-fit model,
finding an equally acceptable fit ($\chi^{2}/\nu = 89.0/91$);
however, in this case, the column density was much lower ($\sim$$10^{22}$ cm$^{-2}$)
while the covering fraction was compatible with the \emph{ASCA} measurement. 
This is very similar to what has been observed in similar sources where the absorption 
changes but the covering fraction remains constant (see, for example, Malizia et al. 2003 for Mrk 6). 
It also explains why the 2--10 keV flux observed by \emph{Swift} XRT, 
\emph{ASCA} and \emph{RXTE} (Revnivtsev et al. 2004)
is highly variable, by about a factor of 2 on
long ($\sim$years) time-scales, with the XRT measurement being in the
middle of the range of the observed values, ($\sim$0.5--1.4)$\times10^{-11}$ erg
cm$^{-2}$ s$^{-1}$.

To conclude, it is important to stress that this source is not Compton 
thick, but more likely an absorbed intermediate Seyfert 1 galaxy, where the 
BLR is observed through the upper layers of the torus. This is further 
confirmed by our diagnostic ratios, which locate the object in a region 
populated by Compton thin Seyfert galaxies (see Table~\ref{Tab7}). 

\subsection{Swift J1009.3--4250}

This \emph{Swift} source (Ajello et al. 2006b) coincides in position with 
ESO 263--13 (= Fairall 427), a face-on spiral galaxy at redshift $z$=0.0333$\pm$0.0003, 
classified as a Seyfert 2 galaxy (Fairall 1983); it is a peculiar source as it has a double
nucleus (Gimeno et al. 2004) and it shows an extended ionized nebulosity (Durret 1989).
This object is also a radio emitter, detected in the SUMSS catalog with a 
36 cm flux of $23.1\pm1.1$ mJy, and it is listed in the extended 2MASS catalog.

According to Ajello et al. (2006b), a model including an absorbed power law plus a 
reflection component ({\sc pexrav} model in XSPEC) provides a good description of 
the combined XRT/BAT spectra, yielding a photon index $\Gamma$ $\sim$2 and a lower 
limit for the absorbing column density of $\sim$$10^{24}$ cm$^{-2}$, again suggesting 
a Compton thick nature.

\begin{figure}
\rotatebox{-90}{\epsscale{0.7} \plotone{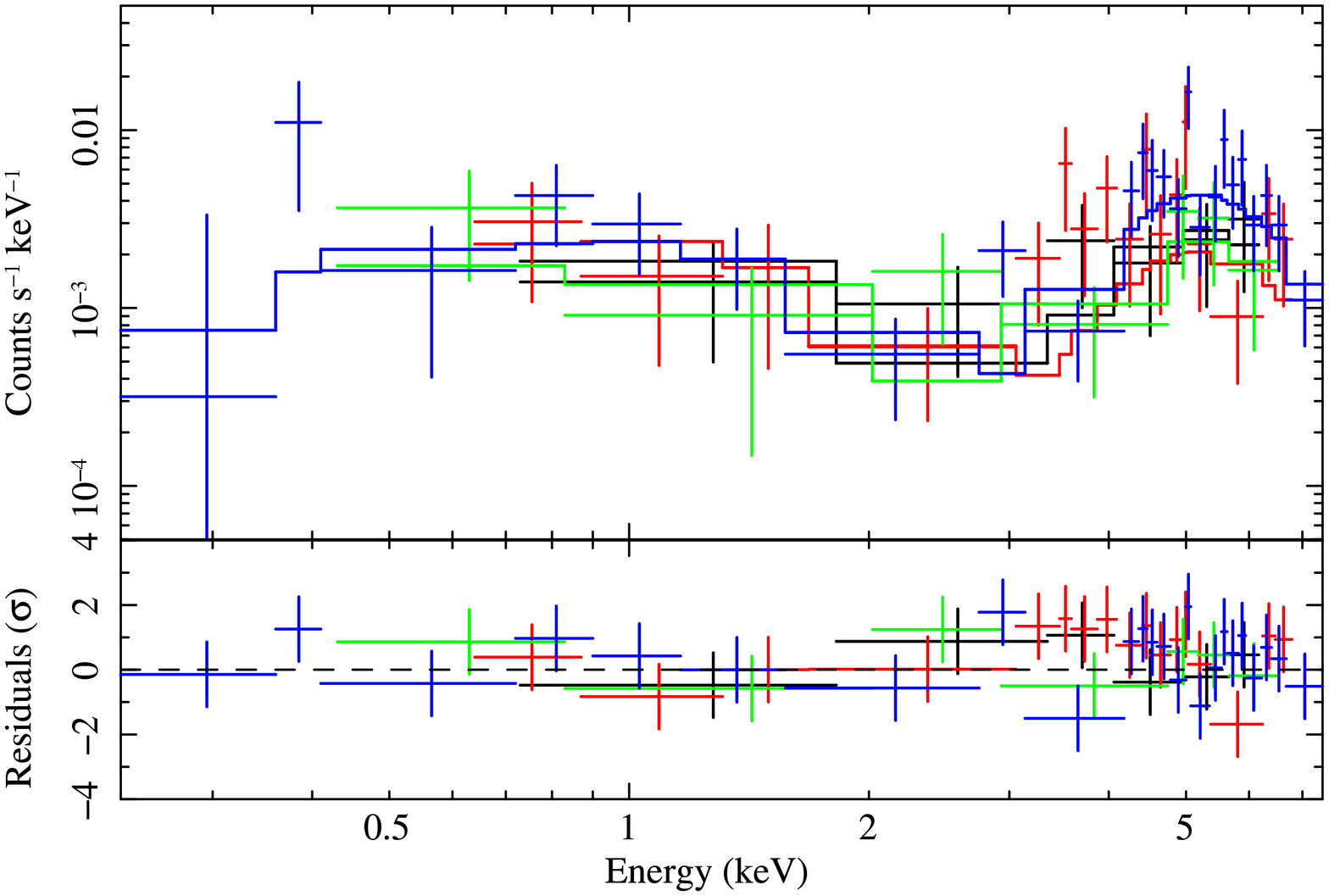}}
\caption{Four XRT spectra of Swift J1009.3--4250 fitted with a primary absorbed power
law component plus a power law, having the same photon index, but absorbed only by Galactic
column density (upper panel); residuals to this model in units of $\sigma$ (lower panel).}
\label{spec1009}
\end{figure}

Our XRT data of Swift J1009.3--4250 are well described by our baseline model
plus another power law, having the same photon index, but absorbed only by the Galactic column 
density; the extra power law component is required 
(at a confidence level $>$ 99.99\% in the combined spectrum and in the range
95.4--99.99\% in each single pointing)  
to account for excess emission observed below 2 keV (see Figure~\ref{spec1009}).
The observed column density is 3--4$\times10^{23}$ cm$^{-2}$, well below the Compton thick regime.
We also found a flux variation of $\sim$50$\%$ among the XRT observations, but no evidence
for spectral changes.

The 6dF optical spectrum of ESO 263--13 shows a flat continuum with
the presence of narrow Balmer, He {\sc ii},
[N {\sc ii}], [S {\sc ii}], [O {\sc i}], 
[O {\sc ii}] and [O {\sc iii}] emissions, with FWHM $\la$ 600 km 
s$^{-1}$ (Figure~\ref{optical}, lower left panel) and at a redshift 
$z=0.0335\pm0.0001$ (Jones et al. 2004), which is consistent
with the result of Fairall (1983). This corresponds to a 
luminosity distance $D_L$ = 159 Mpc and gives a luminosity of
(4.5--8.4)$\times10^{42}$ erg s$^{-1}$ in the 2--10 keV energy band. 
The narrowness of the lines and their relative strengths as seen in the 6dF
spectrum confirm the Seyfert 2 nature of this galaxy.

The H$_\alpha$/H$_\beta$ flux ratio of 4.0 provides an optically dereddened [O {\sc iii}] flux 
of 
$F_{\rm [OIII]} = 2.7\times10^{-13}$ erg cm$^{-2}$ s$^{-1}$ which, when compared 
with the 2--10 keV X-ray flux, implies an AGN in the Compton thin regime (see Table~\ref{Tab7}).
This is confirmed by the $L_{\rm X}$/$L_{\rm HX}$ ratio, which, although low, is 
nevertheless compatible with the observed column
density. We therefore rule out the Compton thick hypothesis put forward
by Ajello et al. (2006b).

\subsection{Swift J1038.8--4942}

This source (Tueller et al. 2005) is likely associated with an extended 
2MASS object (2MASX J10384520--4946531), which is identified with an unclassified
galaxy. No redshift is known, and no radio emission has so far 
been reported from this object. The source is also listed in the 
\emph{ROSAT} All Sky Survey Bright Source Catalog (Voges et al. 1999) and is 
associated with an \emph{IRAS} object.

\begin{figure}
\rotatebox{-90}{\epsscale{0.7} \plotone{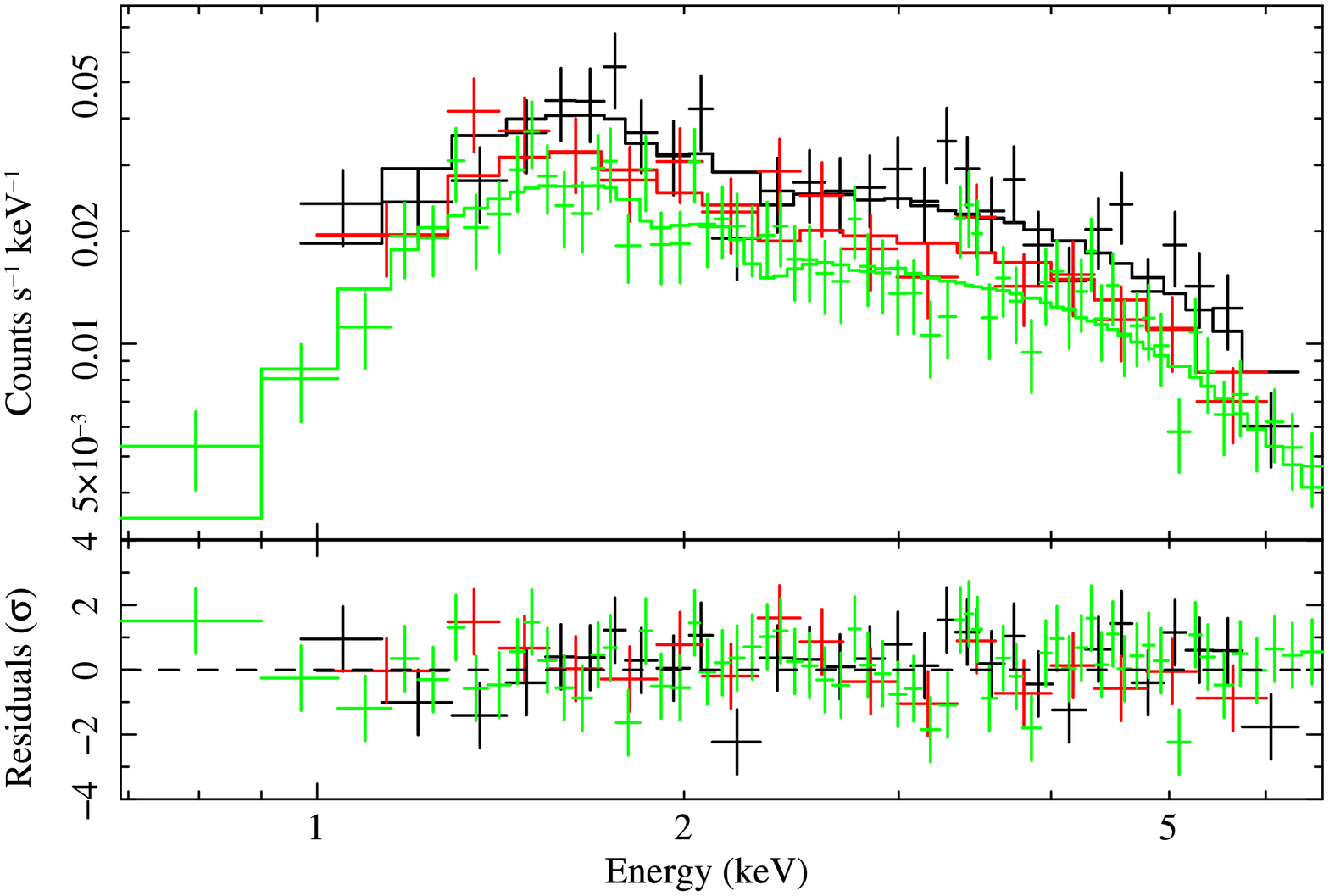}}
\caption{Three XRT spectra of Swift J1038.8--4942 fitted with an absorbed power law 
(upper panel); residuals to this model in units of $\sigma$ (lower panel).}
\label{spec1038}
\end{figure}

Preliminary analysis of 
the XRT data (Tueller et al. 2005) provided information only on the absorption 
($N_{\rm H}$ $\sim$ 10$^{22}$ cm$^{-2}$).
Our analysis indicates either a flat spectrum and lower absorption or a more canonical one with
a similar column density (see Table~\ref{Tab2} and Figure~\ref{spec1038}); 
furthermore, we detect variations in flux of $\sim$10--30$\%$ between observations.

Optical spectroscopy of this object shows broad H$_\alpha$ and H$_\beta$ 
emissions (FWHM$_{\sc H_\beta}$ $\sim$6700 km s$^{-1}$) with a narrow 
component on top; moreover, narrow
[N {\sc ii}], [S {\sc ii}], [Ne {\sc iii}], [O {\sc i}] [O {\sc ii}] and [O 
{\sc iii}] emission lines are detected on a relatively flat continuum 
(Figure~\ref{optical}, lower right panel). The H$_\beta$/[O {\sc iii}] flux 
ratio, 1.7, allows us to classify this galaxy as a Seyfert of type 1.5. From the observed 
wavelengths of narrow features, we determine for the first time 
a redshift $z=0.060\pm0.001$ for this galaxy, which means a luminosity distance 
$D_L$ = 289.3 Mpc. This corresponds to a 2--10 keV luminosity range 
(7.8--11.7)$\times$10$^{43}$ erg s$^{-1}$ and, assuming a 
magnitude $B$ $\sim$16.1 and no local absorption in the optical, an 
absolute $B$-band magnitude M$_B$ $\sim$ $-$23.2.

Given its optical classification, this object is obviously Compton thin, as confirmed by 
our diagnostic ratios; it is, however, another interesting case of an intermediate Seyfert 
with intrinsic absorption.

\subsection{Swift J1200.8+0650}

This BAT source (Kennea et al. 2005) has been associated with PGC 
037894 (also CGCG 041--020 and LEDA 37894), a principal galaxy at 
redshift $z=0.036$, which has not been optically classified yet. The 
source is a weak radio emitter detected in the NVSS and FIRST (Faint 
Images of the Radio Sky at Twenty-centimeters, White et al. 1997) 
surveys with a 20 cm flux in the range 5--6 mJy; like other 
AGNs of our sample, it is an \emph{IRAS} faint source
and a 2MASS extended object.

\begin{figure}
\rotatebox{-90}{\epsscale{0.7} \plotone{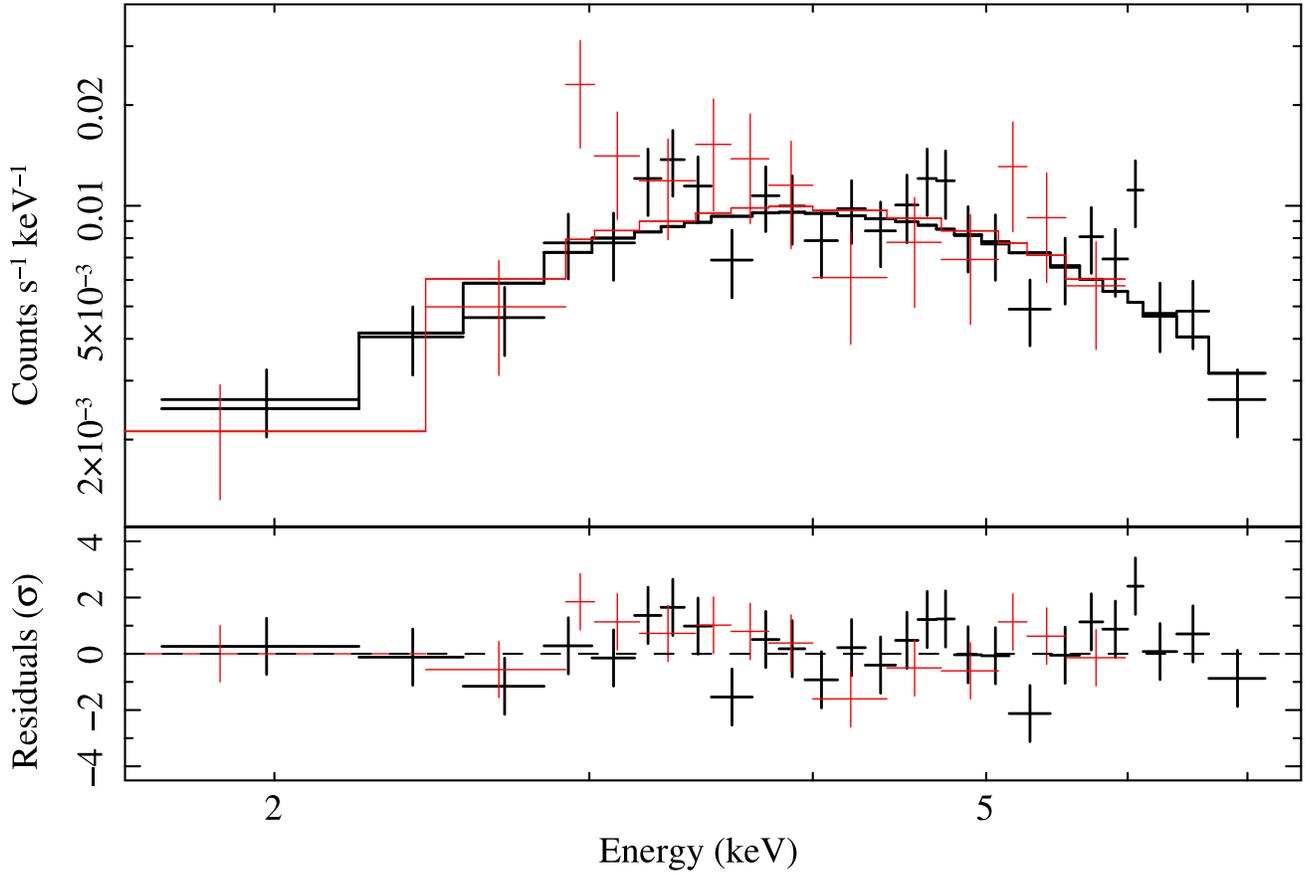}}
\caption{Two XRT spectra of Swift J1200.8+0650 fitted with an absorbed power law (upper panel);
residuals to this model in units of $\sigma$ (lower panel).} 
\label{spec1200}
\end{figure}

In this case, no X-ray emission has so far been reported.
Individual or combined X-ray data provide an acceptable fit with an absorbed power law having 
a column density of (6--8)$\times10^{22}$ cm$^{-2}$ and a photon index in the range 1.3--1.8
(see Table~\ref{Tab2} and Figure~\ref{spec1200}).
 
The addition of a narrow Gaussian Fe K$\alpha$ line 
in the XRT spectrum of the first observation (thicker crosses in Figure~\ref{spec1200})
does not significantly improve the quality of the fit ($\Delta\chi^{2}=7.1$ for 2 
degrees of freedom or $>$ 95.4\% confidence level), but returns a
line energy of E $\sim$6.3 keV and an $EW$ of $\sim$266 eV, 
much in line with other AGN measurements of this feature. It is likely that a longer dedicated 
exposure could provide further insight on the presence of this feature.

The optical spectrum of PGC 037894 shows H$_\alpha$, H$_\beta$, [N 
{\sc ii}], [S {\sc ii}] and [O {\sc iii}] narrow (FWHM $\la$ 850 km 
s$^{-1}$) emission lines over a flat continuum (Figure~\ref{optical1}, upper left panel). 
The features are at redshift $z=0.035\pm0.001$, consistent with Grogin, 
Geller \& Huchra (1998), which corresponds to a luminosity distance $D_L$ 
= 165.8 Mpc and gives a $1.9\times10^{43}$ erg s$^{-1}$ 
2--10 keV luminosity for this source. Again, the line shapes
and ratios indicate that this is a Seyfert 2 galaxy.

The H$_\alpha$/H$_\beta$ flux ratio 3$\sigma$ upper limit ($>$ 4.7) 
implies  a dereddened 
[O {\sc iii}] flux $F_{\rm [OIII]} > 2.9\times10^{-14}$ erg cm$^{-2}$ 
s$^{-1}$. The diagnostic ratios, and also the possible iron line $EW$ value, place this 
source well in the Compton thin regime, indicating that is a classical type 2 AGN.

\subsection{Swift J1238.9--2720}

This source (Tueller et al. 2005) is associated with the 
optically unclassified galaxy ESO 506--G027, at $z=0.025$ (Da Costa et 
al. 1998). This is a radio emitter with a 20 cm flux of $73.7\pm2.3$ 
mJy in the NVSS survey. This galaxy is also an \emph{IRAS} faint source
and it is an extended 2MASS object.
 
\begin{figure}
\rotatebox{-90}{\epsscale{0.7} \plotone{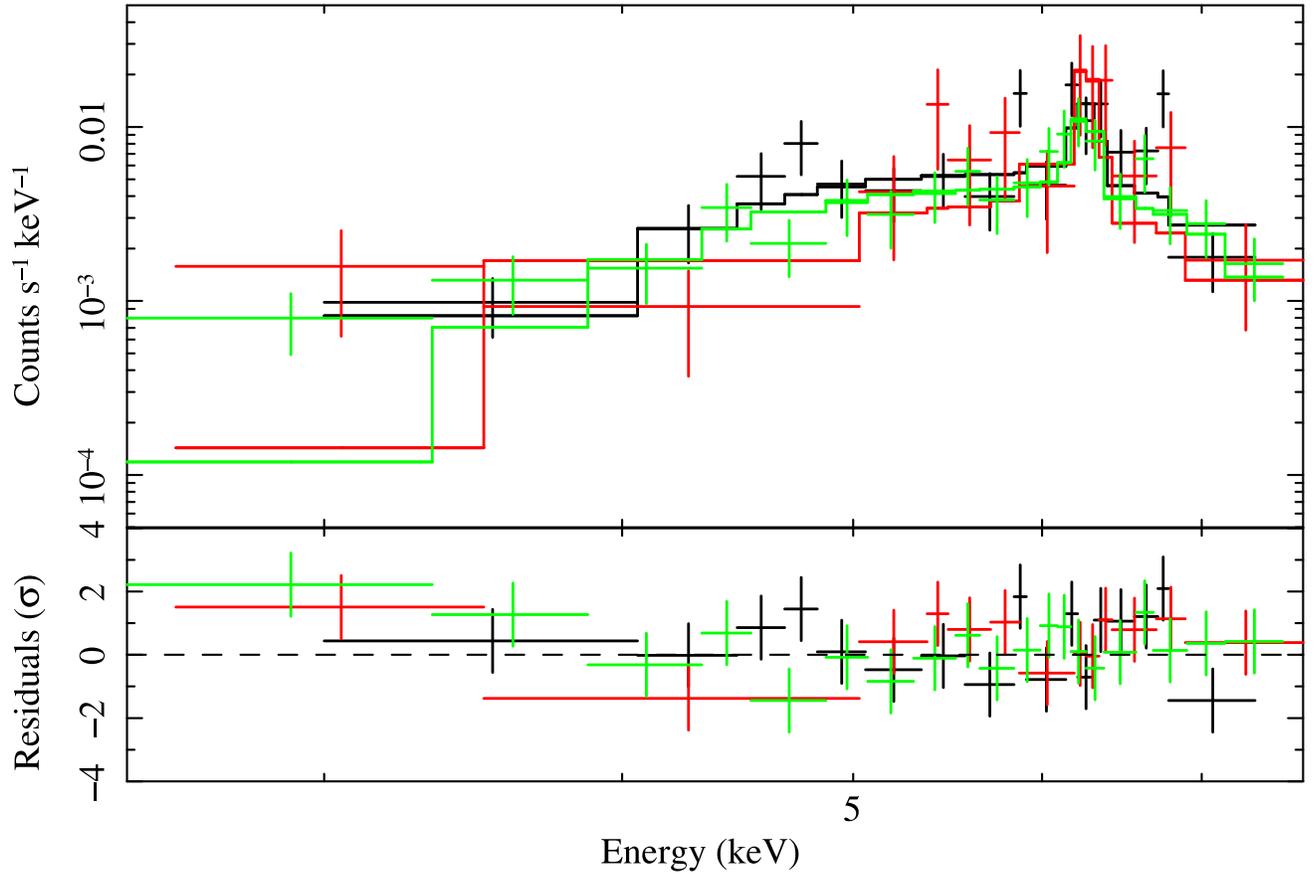}}
\caption{Three XRT spectra of Swift J1238.9--2720 fitted with an 
absorbed power law plus a narrow Gaussian Fe K$\alpha$ line 
(upper panel); residuals to this model in units of $\sigma$ (lower panel).}
\label{spec1238}
\end{figure}

\begin{figure}
\rotatebox{-90}{\epsscale{0.7}\plotone{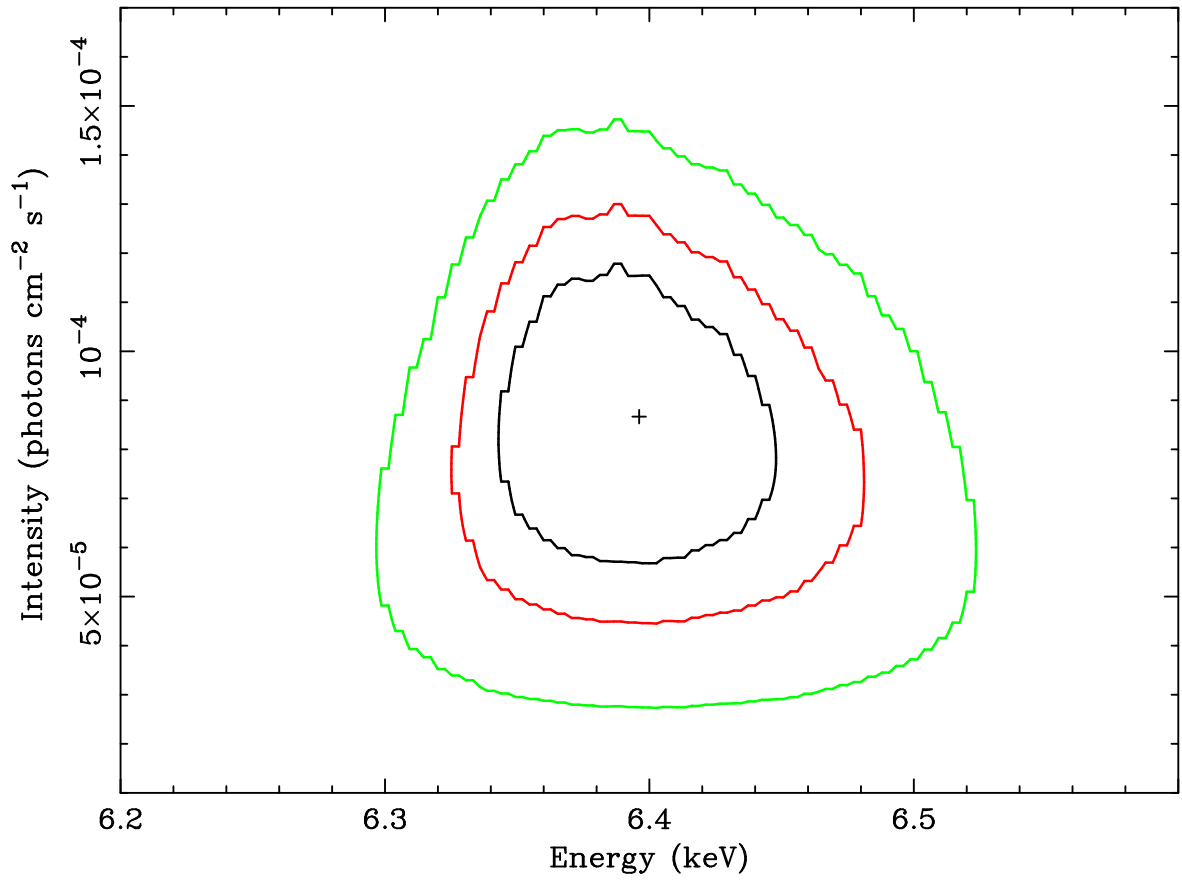}}
\caption{Confidence contours (at 68\%, 90\%, and 99\% confidence level) 
of the iron line energy versus intensity for the best-fit model
of Swift J1238.9--2720.} 
\label{spec1238line}
\end{figure}

A preliminary analysis of the XRT data indicated that it might be a 
Compton thick object showing strong absorption ($N_{\rm H}$ 
$\sim$10$^{23}$ cm$^{-2}$) and an iron line with  
$EW$ of about 500 eV (Tueller et al. 2005).
The data indicate a change in flux ($\sim$20\%) between the first and 
second observations on a time-scale of a few months (see Table~\ref{Tab3});
this evidence casts some doubts on the Compton thick nature of the source as the reflection 
component, if due to the torus, is expected to provide flux changes on longer time-scales.
Individual and combined spectra are not of sufficient
quality to allow a constraint on the photon index which was therefore fixed 
to the canonical value of 1.8; the observed column density is $\sim$10$^{23}$ cm$^{-2}$ 
and a narrow iron line is detected with $>$ 99.73\% 
confidence ($\Delta\chi^{2}=21.1$ for 5 degrees of freedom) (see Figure~\ref{spec1238}).
The two-dimensional iso--$\chi^{2}$ contour plot of line intensity versus 
energy is shown in Figure~\ref{spec1238line}. 
The line has an $EW$ of 500 eV which does not necessarily imply that we are dealing
with a Compton thick AGN, as a similar value is also compatible with the observed  
column density (Turner et al. 1997).

The optical spectrum (Figure~\ref{optical1}, upper right panel) of this source may  
help in assessing its Compton nature. The presence of H$_\alpha$, H$_\beta$, 
[N {\sc ii}] and [O {\sc iii}] narrow (FWHM $\la$ 1000 km s$^{-1}$) 
emission lines detected on a spiral galaxy continuum and the overall 
spectral appearance indicate that this source is most likely a Seyfert 2 galaxy. 
All emission features are at redshift $z=0.024\pm0.001$, consistent 
with Da Costa et al. (1998). This implies a luminosity distance $D_L$ = 
112.7 Mpc and a (6.4--8.2)$\times$10$^{42}$ erg s$^{-1}$ 2--10 keV 
luminosity range for this galaxy.

The H$_\alpha$/H$_\beta$ flux ratio 3$\sigma$ upper limit of $>$ 23, 
gives a dereddened [O {\sc iii}] flux greater than $1.2\times10^{-11}$ erg. 

The flux ratios provide contradictory results as the $L_{\rm X}$/$L_{\rm [OIII]}$ 
indicates a Compton thick nature, while the $L_{\rm X}$/$L_{\rm IR}$ and 
$L_{\rm X}$/$L_{\rm HX}$ provide evidence for a thin behavior.
However, the observed column density is sufficient to explain the low values of the first  
ratio: very likely, this is a borderline object in between the thin and thick Compton regimes.

\subsection{Swift J1930.5+3414}

This source has a counterpart classified as a galaxy 
in the 2MASS extended catalog (2MASX J19301380+3410495) (Kennea et al. 2005).
The source is possibly detected as a radio source located $7^{\prime \prime}$ away
(NVSS J193013+341047) by the NVSS survey, with a 20 cm flux of $3.9\pm0.5$ mJy
and it is also likely an \emph{IRAS} source. It has recently been classified as 
a Seyfert 1 galaxy at a redshift $z=0.0629$ (Halpern 2006) but no X-ray data 
have so far been reported.
 
\begin{figure}
\rotatebox{-90}{\epsscale{0.7} \plotone{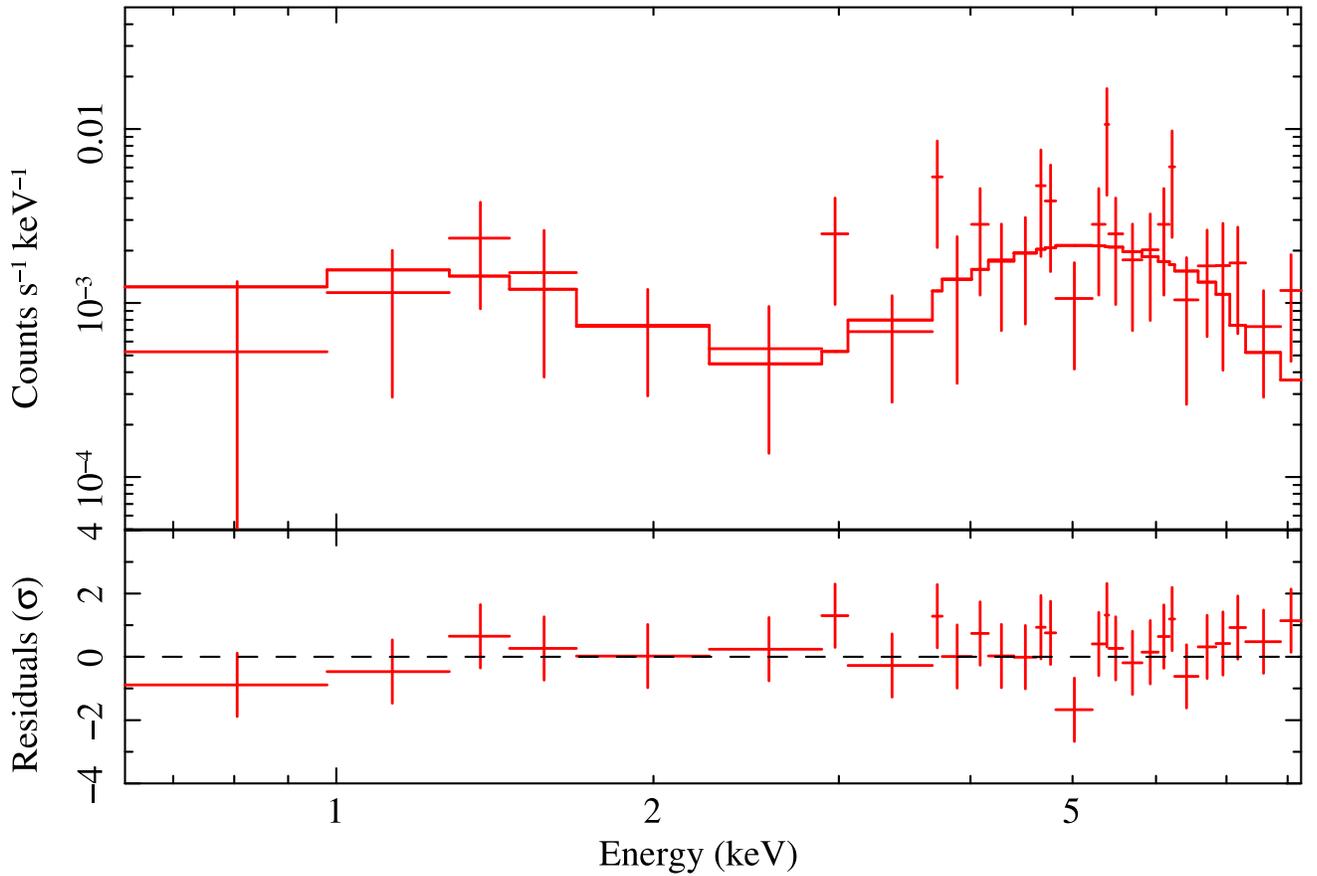}}
\caption{XRT spectrum (obs 2) of Swift J1930.5+3414 fitted with an absorbed power law 
component plus a power law, having the same photon index, but absorbed only by Galactic
column density (upper panel); residuals to this model in units of $\sigma$ (lower panel).}
\label{spec1930}
\end{figure}

The X-ray spectrum is well described (see Fig~\ref{spec1930}) by an absorbed power law with 
photon index frozen at 1.8 plus a power law having the same photon index but only 
absorbed by the Galactic column density. This extra component is required by the data with a
significance $> 99.73\%$ and accounts for excess emission below 2 keV (see Table~\ref{Tab3}).   
We estimate an intrinsic column density of $\sim$$3\times10^{23}$ cm$^{-2}$, 
a value at odds with the Seyfert 1 classification.

In the spectrum of the optical counterpart of Swift J1930.5+3414,
we confirm the emission features reported by Halpern (2006), i.e., 
broad (FWHM $\sim$5800 km s$^{-1}$)
Balmer lines with a narrow component on top and with an overall 
skewed profile, as well as a prominent narrow (FWHM $\sim$900 km s$^{-1}$) 
[O {\sc iii}] line along with narrow [S {\sc ii}], [O {\sc ii}],
[Ne {\sc iii}] emissions and other weaker features, all superimposed on a 
flat continuum 
(Figure~\ref{optical1}, central left panel). The narrow emission lines are at 
redshift $z=0.063\pm0.001$, in agreement with Halpern (2006).
This corresponds to a luminosity distance $D_L$ = 303.9 Mpc and gives a 
2--10 keV luminosity of $2.0\times10^{43}$ erg s$^{-1}$ for this source. 
Likewise, assuming a magnitude $B$ $\sim$17.0, one gets
an absolute $B$-band magnitude M$_B$ $\sim$ $-$21.2.
This should however be considered as a conservative upper limit 
as no absorption local to the host galaxy was accounted for.

The H$_\beta$/[O {\sc iii}] flux ratio, 0.34, allows us to 
revise the Seyfert 1 AGN classification given by Halpern (2006) and to
classify this object as an intermediate Seyfert of type 1.5--1.8;
the absorption detected suggests that a higher Seyfert classification 
may be more appropriate.
Furthermore, the diagnostic ratios indicate that this AGN operates in the
Compton thin regime.

\subsection{Swift J1933.9+3258}

This source (Grupe et al. 2006) was first discovered as a bright X-ray source during 
the \emph{ROSAT} All Sky Survey as 1RXS J193347.6+325422 (Voges et al. 1999).
The object has a counterpart in the 2MASS and \emph{IRAS} catalogs 
and it is also listed in the NVSS survey (NVSS J193347+325426) with a 20 cm 
flux of $4.0\pm0.5$ mJy. 
The \emph{Swift} BAT detection triggered an optical spectroscopic observation, which
allowed the classification of the source as a 
Seyfert 1.2 galaxy at $z=0.0580\pm0.0001$ (Torres et al. 2006).
Preliminary analysis of the XRT data (Grupe et al. 2006) provided a best-fit model 
of a broken power law with a soft photon index of $\sim$3, a hard photon index of 
$\sim$2, $E_{\rm B}$ $\sim$1.3 keV and an absorption column density $N_{\rm H}$ 
$\sim$$3.4\times10^{21}$ cm$^{-2}$, which is in agreement with the value found with \emph{ROSAT} data.

\begin{figure}
\rotatebox{-90}{\epsscale{0.7} \plotone{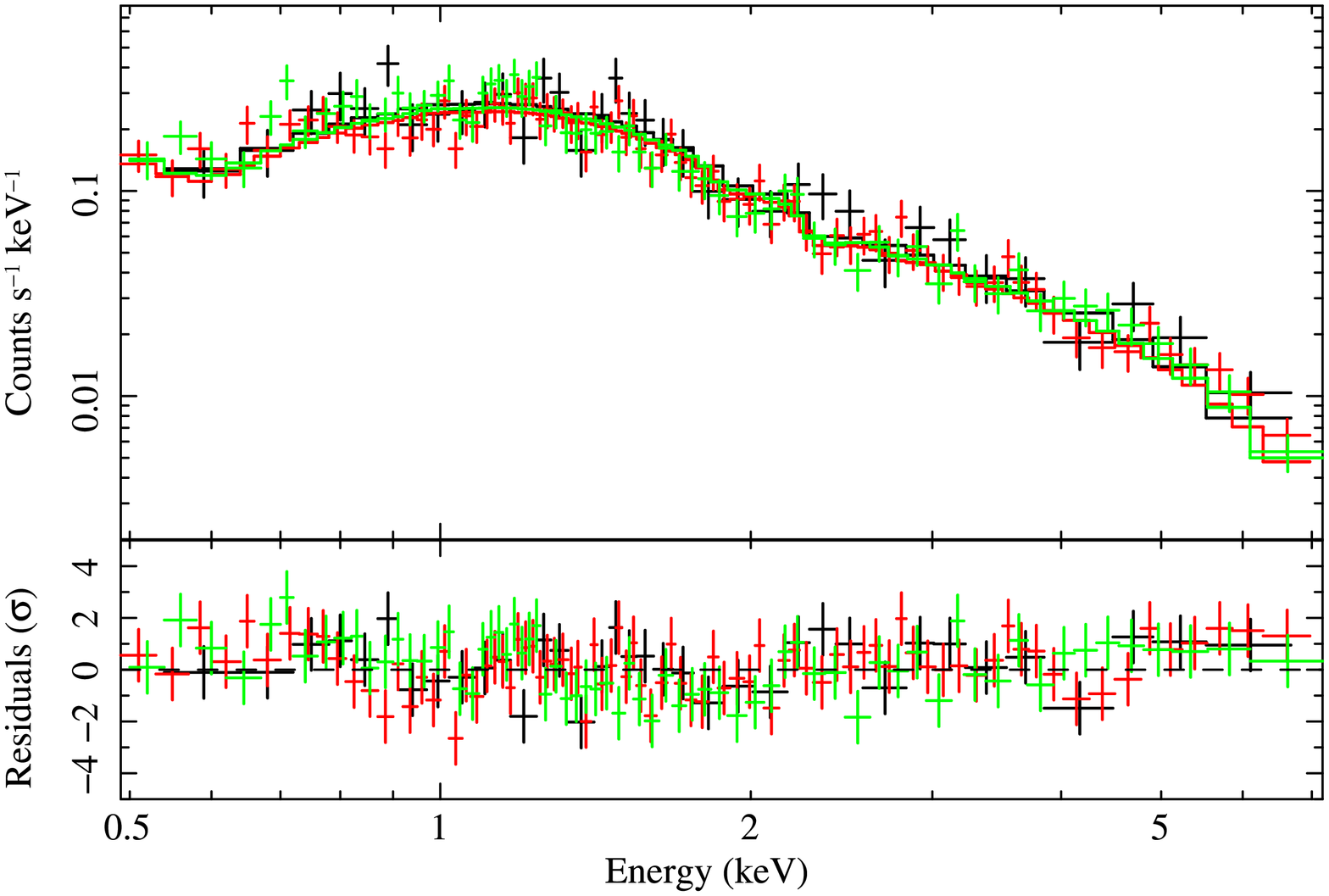}}
\caption{Three XRT spectra of Swift J1933.9+3258 fitted with an
absorbed power law (upper panel); residuals to this model in units of $\sigma$ (lower
panel).}
\label{spec1933}
\end{figure}

Our XRT analysis, although consistent with a broken power law description,
is also compatible with a simple power law model (see Fig~\ref{spec1933}) with $\Gamma$ $\sim$2 
(see Table \ref{Tab2}); in either case, we do not find evidence for absorption 
($N_{\rm H}<6\times10^{21}$ cm$^{-2}$) in excess of the Galactic value. 

The optical spectra we acquired show broad He {\sc i}, He {\sc ii}, 
Fe {\sc i}, and Balmer emissions (FWHM$_{\sc H_\beta}$ $\sim$3800 km 
s$^{-1}$) at least up to H$_\zeta$, as well as narrow (FWHM $\sim$800 km 
s$^{-1}$) [O {\sc iii}] and [Ne {\sc iii}] emission lines (Figure~\ref{optical}, 
central right panel) on a blue continuum. The spectral appearance, together with 
a H$_\beta$/[O {\sc iii}] flux ratio ($\sim$3.5), implies a Seyfert 1.2 
classification for this galaxy, in agreement with Torres et al. (2006); 
the redshift inferred from the optical spectrum is $z=0.063\pm0.001$, also in agreement with 
Torres et al. (2006).

This gives a luminosity distance $D_L$ = 284.3 Mpc to the source, which
corresponds to a luminosity of $7.6\times10^{43}$ erg s$^{-1}$ 
in the 2--10 keV band and an absolute $B$-band magnitude 
M$_B$ $\sim$ $-$24.5 (assuming $B$ $\sim$13.9).
The overall picture, including the value of the diagnostic ratios, is that of a ``canonical" 
Seyfert galaxy of type 1. 

\subsection{XSS J12303--4232}

This source is listed in the \emph{RXTE} Slew Survey (XSS, Revnivtsev 
et al. 2004); it is associated with a bright \emph{ROSAT} object
localized with sufficient accuracy to allow its 
identification with an \emph{IRAS} object (F12295--4201) that is still unclassified.
A previous analysis of the XRT data suggested that the spectrum was 
typical of a type 1 AGN and that it contained a redshifted ($z$ $\sim$0.1) iron line 
(Revnivtsev et al. 2006).

\begin{figure}
\rotatebox{-90}{\epsscale{0.7} \plotone{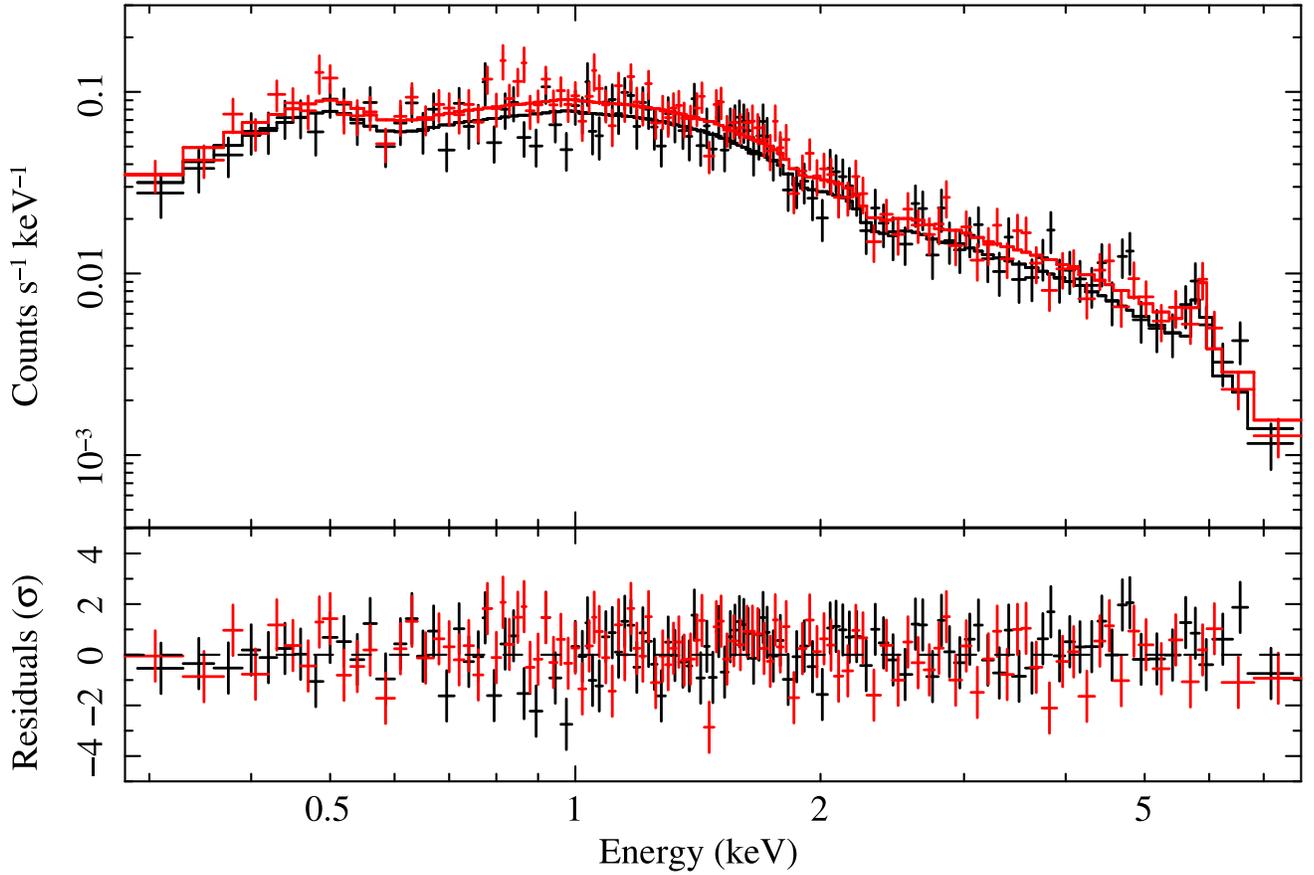}}
\caption{Two XRT spectra of XSS J12303--4232 
fitted with an absorbed power law plus a narrow Gaussian Fe K$\alpha$ line
(upper panel); residuals to this model in units of $\sigma$ (lower panel).}
\label{spec12303}
\end{figure}

Our own analysis of individual and combined XRT data confirms the presence of 
a canonical AGN spectrum (see Table~\ref{Tab2} and Figure~\ref{spec12303}); 
an iron line is required by the data with $>$ 99\% 
confidence level ($\Delta\chi^{2}=12$ for 2 degrees of freedom).

The \emph{RXTE} flux of $\sim$$5\times10^{-12}$ erg cm$^{-2}$ s$^{-1}$ in the 3--8 keV energy
band is fully compatible with the two XRT measurements.

Optical spectroscopy of the object at the XRT position shows broad Balmer, 
He {\sc i}, He {\sc ii} and Fe {\sc i} emissions (FWHM$_{\sc H_\beta}$
$\sim$5800 km s$^{-1}$) plus narrow [O {\sc i}], [O {\sc ii}] and [O {\sc 
iii}], [N {\sc ii}], [S {\sc ii}], [Ne {\sc iii}] emission lines, 
detected on a flat spectral continuum (Figure~\ref{optical1}, lower left panel); 
H$_\alpha$ and H$_\beta$ also show a narrow component on top of the broad 
one. The H$_\beta$/[O {\sc iii}] flux ratio, 1.1, allows us to classify 
this object as a Seyfert 1.5. From the observed wavelengths of the narrow 
features, we determine for the first time a firm redshift for this galaxy at 
$z=0.100\pm0.001$, definitely compatible with the value inferred by X-ray analysis of the 
iron line. The source luminosity distance is $D_L$ = 495.7 Mpc, providing a  
(9.1--15.3)$\times$10$^{43}$ erg s$^{-1}$ luminosity 
range and, assuming $B$ $\sim$14.3 and no local absorption, an absolute 
magnitude M$_B$ $\sim$ $-$24.6.

As expected, the diagnostic ratios further confirm that this source
is in the Compton thin regime.
 
\section{Conclusions}

We have presented optical classification and X-ray characterization of 12
hard X-ray selected objects: 11 are associated with active galaxies                   
and only one turned out to be an Intermediate Polar (magnetic) CV at $\sim$170 pc from
Earth (see Appendix). 
Optical spectroscopy allowed, for the first time, the classification of
8 objects and a distance determination for 3 of them.
For another object a more refined optical classification than that available in the literature is
also provided.
In the 3 remaining objects, optical spectroscopy provides a characterization
of the source closer in time to the X-ray measurement.
Of this AGN sample, 6 objects are Seyfert 2 galaxies and 5 are Seyferts of 
intermediate type 1.2--1.8; $\sim$70$\%$ of the sample is absorbed with 
$N_{\rm H} > 10^{22}$ cm$^{-2}$.
At least 2 type 1--1.5 objects show absorption in excess of the Galactic value; particularly 
interesting is the case of Swift J0918.5+1618 (Mrk 704) in which the absorption 
is complex and may be linked to the polar scattered nature of the source.
At most 3 objects (Swift J0444.1+2813, Swift J0601.9--8636 and Swift J1238.9--2720)
could be Compton thick.
However, in the case of Swift J0444.1+2813 and Swift J1238.9--2720 the diagnostic ratios
provide controversial results: while the $L_{\rm X}$/$L_{\rm OIII}$ ratio indicates a 
Compton thick object, two other ratios suggest a Compton thin nature.
Both the upper limit on the iron line $EW$ and the short-term variability argue
against a Compton thick nature for Swift J0444.1+2813.
In the case of Swift J1238.9--2720, the X-ray data analysis indicates 
the presence of 
intrinsic absorption with $N_{\rm H}$ $\sim$$5\times10^{23}$ cm$^{-2}$
and of a 6.4 keV iron line with an $EW$ $\sim$500 eV, and suggests that we are dealing with 
a borderline source, heavily absorbed, but not truly Compton thick. 
The third source, Swift J0601.9--8636 is probably the best candidate Compton thick object in the
entire sample.
Overall, our findings emphasize the need for 
hard X-ray surveys to discover and properly sample the population of absorbed AGNs.

\section{Acknowledgements}
We thank Stefano Bernabei and Roberto Gualandi for the assistance at the telescope 
in Loiano, and Claudio Aguilera and Arturo Gomez for their support during observations at CTIO.
This research has made use of data obtained from SIMBAD database operated at CDS,
Strasbourg, France; the High Energy Astrophysics Science Archive Research Center (HEASARC),
provided by NASA's Goddard Space Flight Center; the USNO-A2.0 and 2MASS catalogs;
the NASA/IPAC Extragalactic Database (NED); the ASI Scientific Data Center; and 
the HyperLeda catalog operated at the Observatoire de Lyon, France. 
This research has been supported by ASI under contracts I/008/07/0 and I/023/05.

\clearpage

\appendix
\section{Swift J0732.5--1331}

This is the only source in our list which is not extragalactic. It 
coincides in position with the Bright \emph{ROSAT} All Sky Survey source (RASS) 1RXS 
J073237.6--133113 (Voges et al. 1999). 

A preliminary analysis of the XRT 
spectrum (Ajello et al. 2006a) indicates that either a power law with 
photon index $\Gamma=1.1$, or a Raymond-Smith model (Raymond \& Smith 1977) with a 
temperature of 60 keV are a good fit to the data; in both cases, the hydrogen column 
density is below 10$^{19}$ cm$^{-2}$, which implies a rather small 
distance from Earth.
A reanalysis of the data by Wheatley, Marsh \& Clarkson (2006) showed that a partial covering 
absorption model allows the fitted temperature to drop to a value of $\sim$20 keV.

Our own analysis of the XRT data are in agreement with the Wheatley, Marsh \& Clarkson (2006)
results, providing a best-fit model of a partially covering absorber ($N_{\rm H}$ $\sim$$10^{23}$ 
cm$^{-2}$ and $C_{\rm f}$ $\sim$60\%)
plus a thermal component ({\sc mekal} model in XSPEC, see Mewe, Gronenschild \& van der Oord 1985) 
having an average $kT$ 
of $\sim$15 keV (see Figure~\ref{spec0732} and Table~\ref{Tab3}). 
We therefore confirm previous findings both for individual observations and for
the average spectrum (see Table~\ref{Tab3}).  

Masetti et al. (2006e) stressed that the \emph{Swift} XRT error box 
encompasses two objects; the brighter one (with $R$ $\sim$12.4) has the 
spectrum of a normal G/K-type Galactic star. The fainter one (with $R$ 
$\sim$14.2), located $\sim$10$^{\prime \prime}$ southeast with respect to the 
brighter object, is instead the true optical counterpart: spectroscopy 
reveals H$_\alpha$, H$_\beta$, He {\sc i} $\lambda$5875 and He {\sc ii} 
$\lambda$4686 in emission at redshift zero, superimposed on a very blue 
and otherwise featureless continuum (Figure~\ref{opt0732}). 
These signatures are typical of the accretion disk of a Galactic X-ray 
binary.

Marsh et al. (2006) subsequently found that the counterpart proposed by 
Masetti et al. (2006e) resolves into two stars approximately 1$^{\prime \prime}$.8  
apart and lying along a North-East/South-West axis, and that the actual 
counterpart is the northeastern star of the close pair. This star is also 
approximately half as bright as its companion in the g' optical filter.

After this report, optical and X-ray follow-up observations revealed
a pulsation period (512.42 s, Patterson et al. 2006; Wheately, Marsh \& Clarkson (2006)) 
and an orbital periodicity (0.2335 d, Thorstensen et al. 2006), which suggest that
this object could be classified as an Intermediate Polar (IP) magnetic Cataclysmic Variable
(CV).

\begin{figure}
\rotatebox{-90}{\epsscale{0.38}\plotone{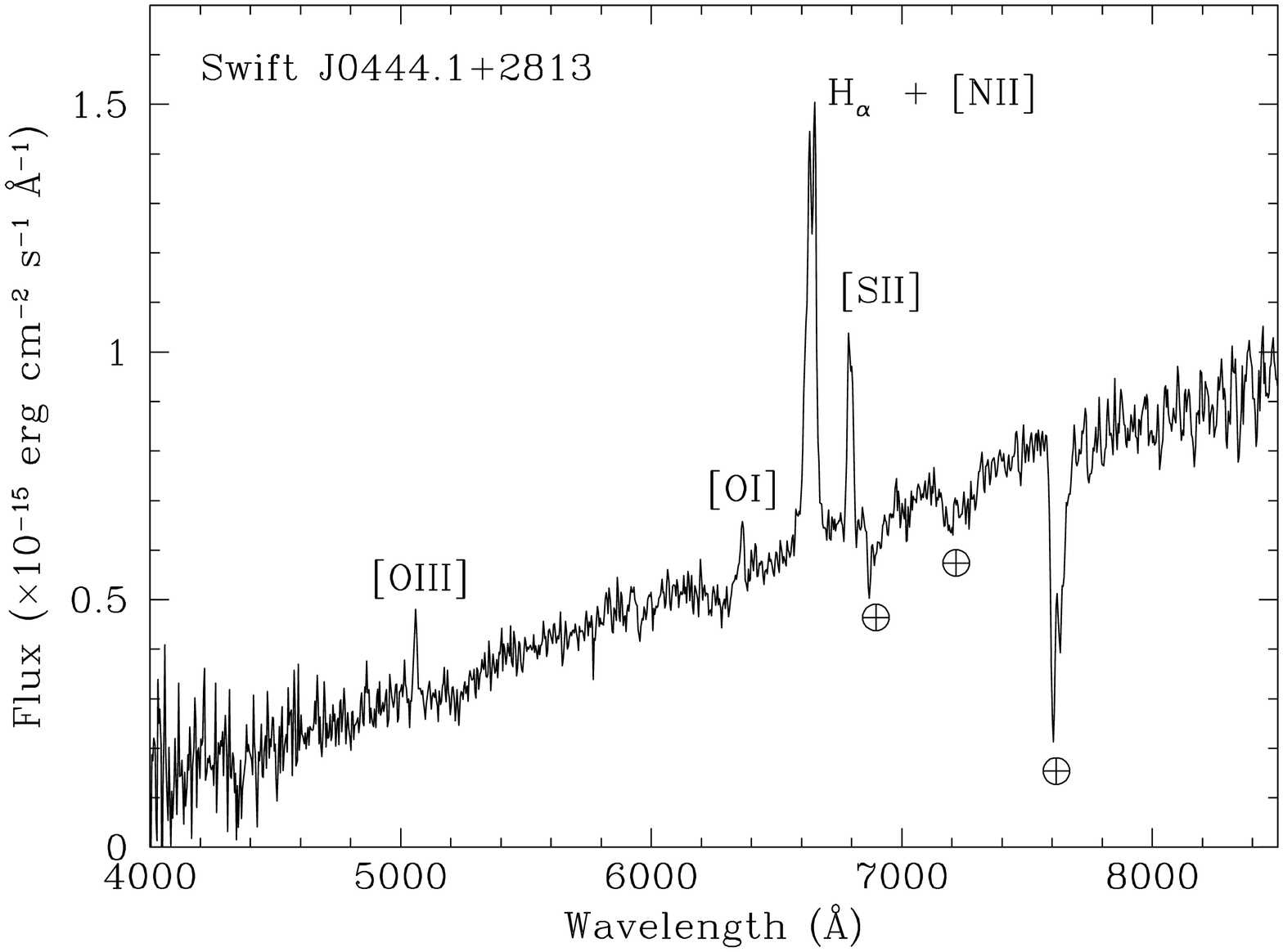}}
\rotatebox{-90}{\epsscale{0.44}\plotone{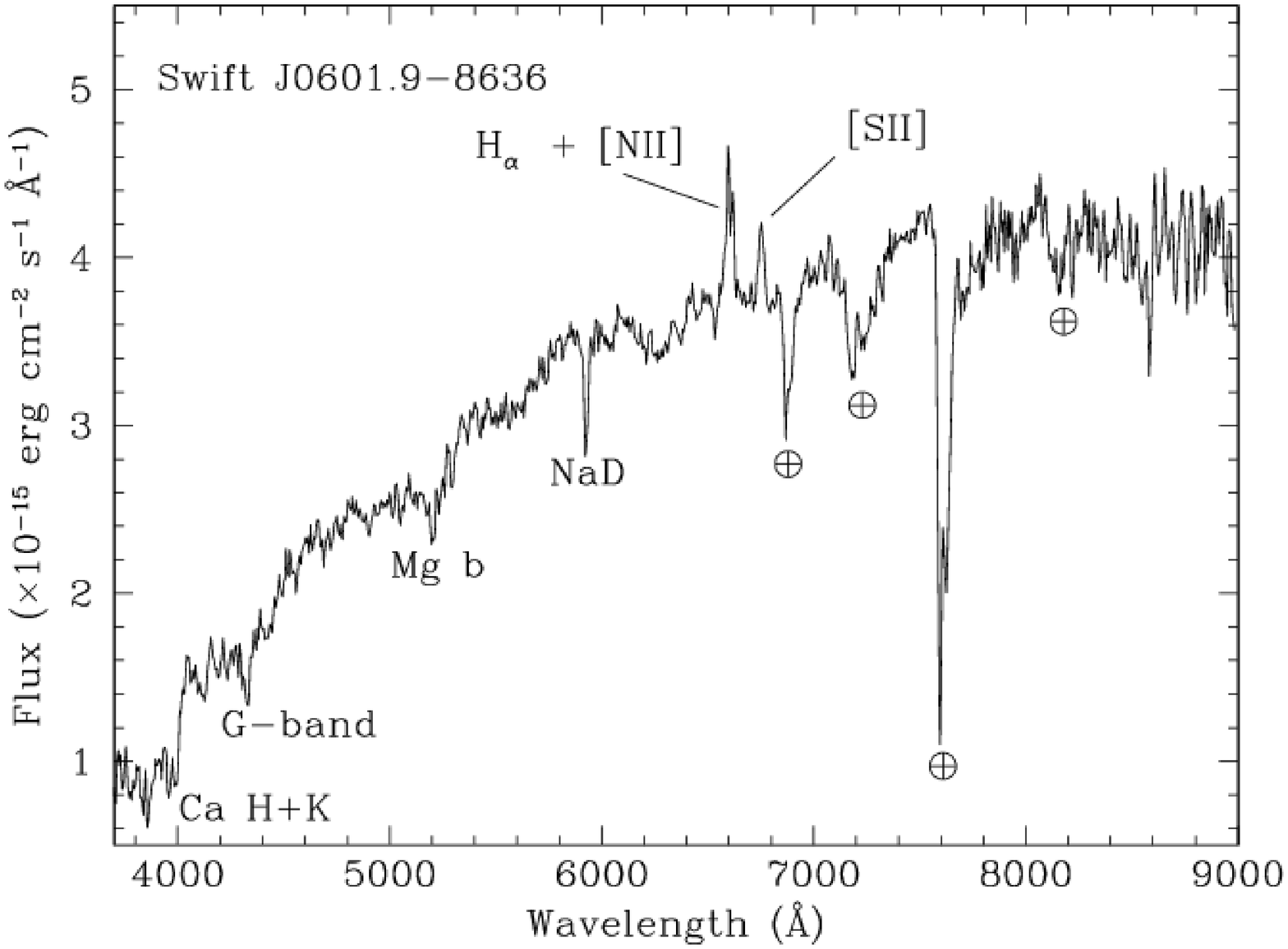}}

\rotatebox{-90}{\epsscale{0.4}\plotone{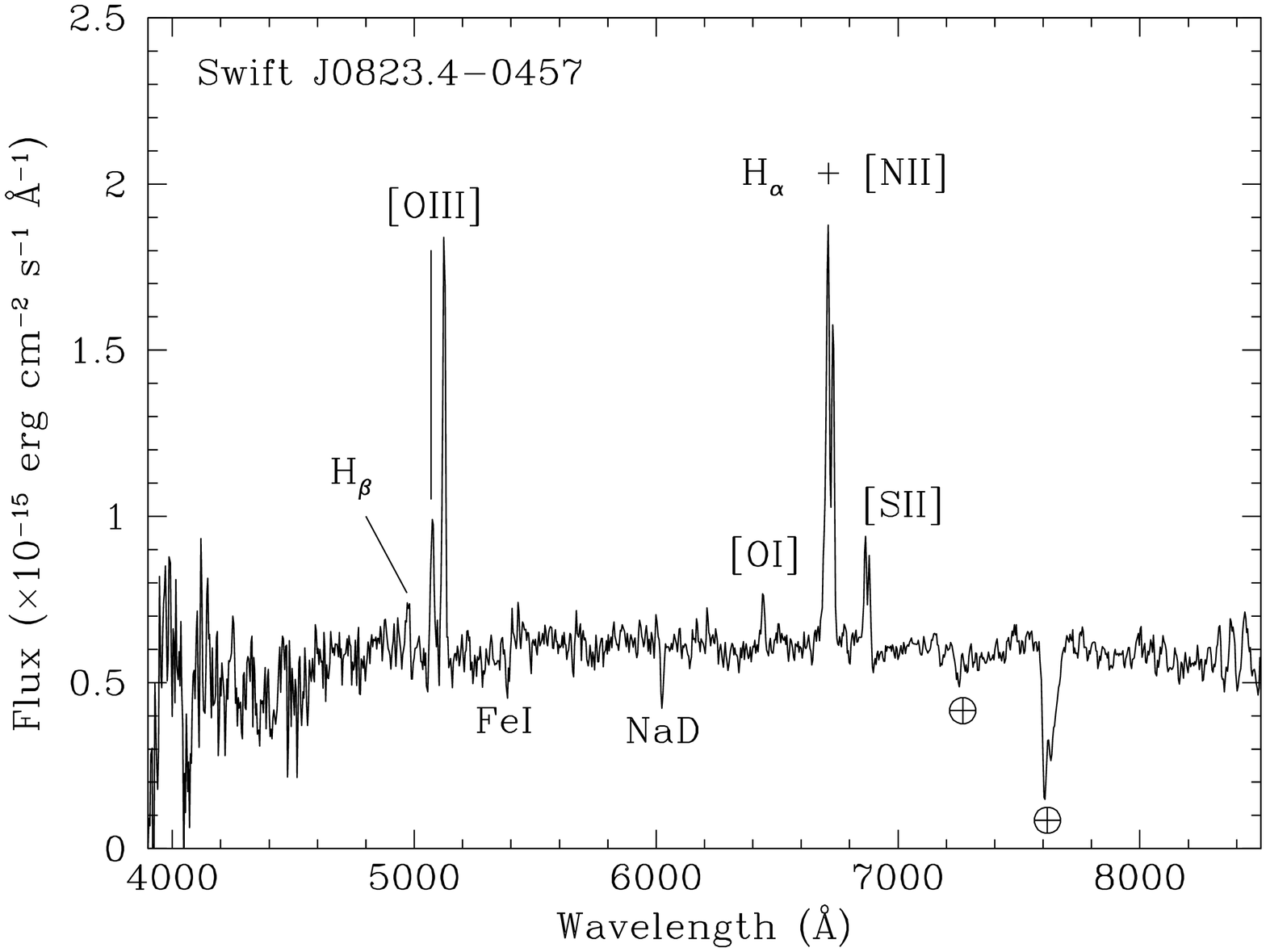}}
\rotatebox{-90}{\epsscale{0.4}\plotone{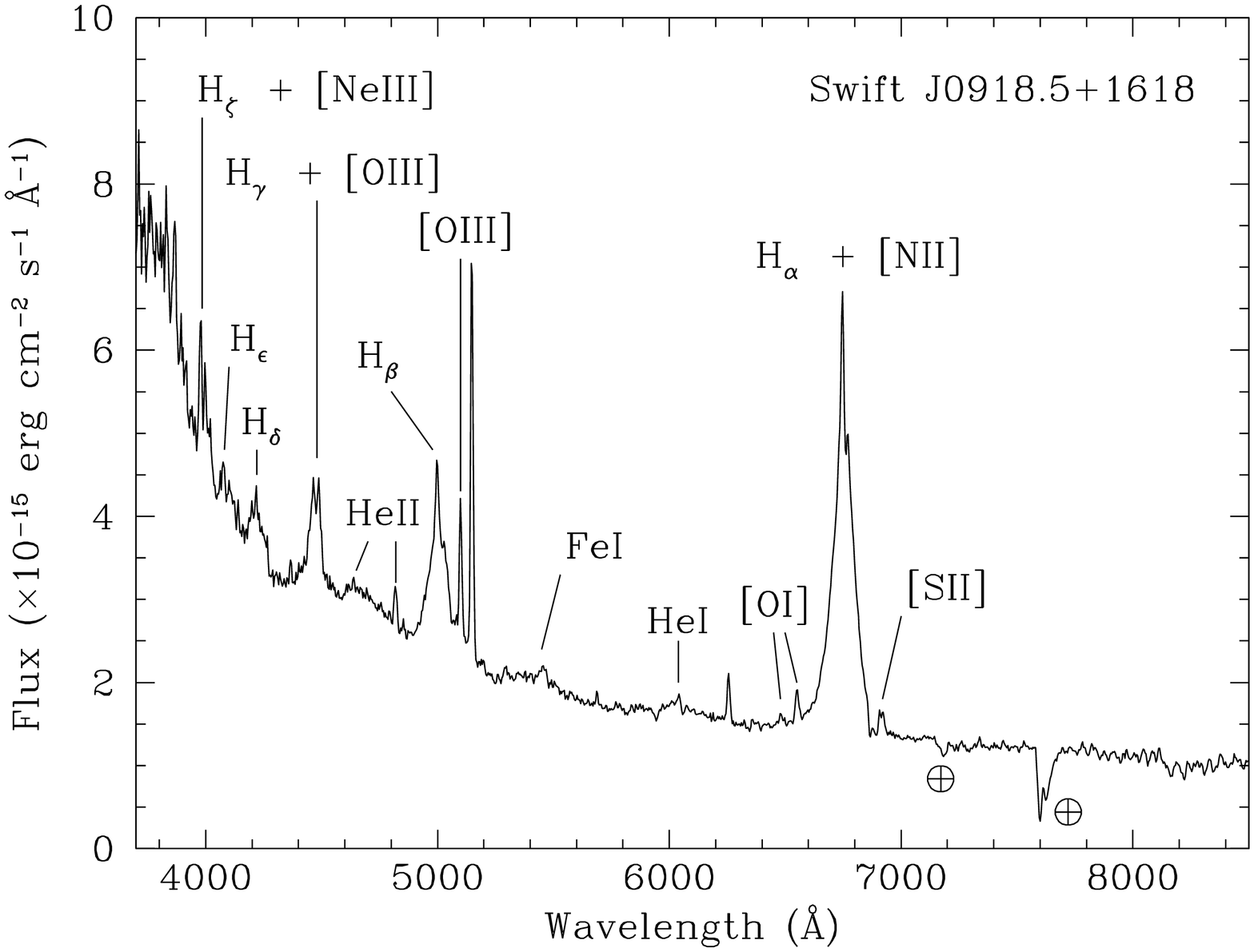}}

\rotatebox{-90}{\epsscale{0.4}\plotone{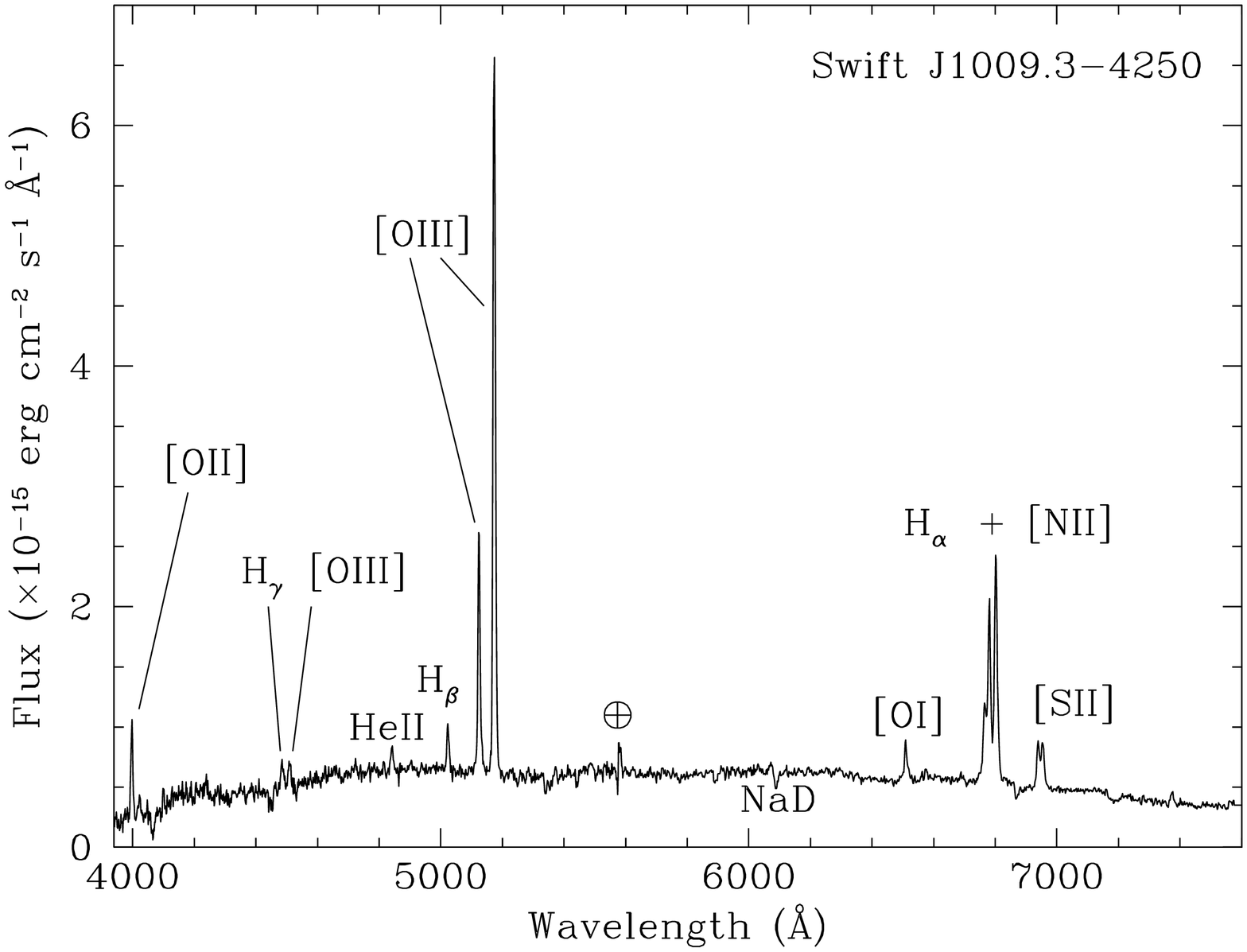}}
\rotatebox{-90}{\epsscale{0.4}\plotone{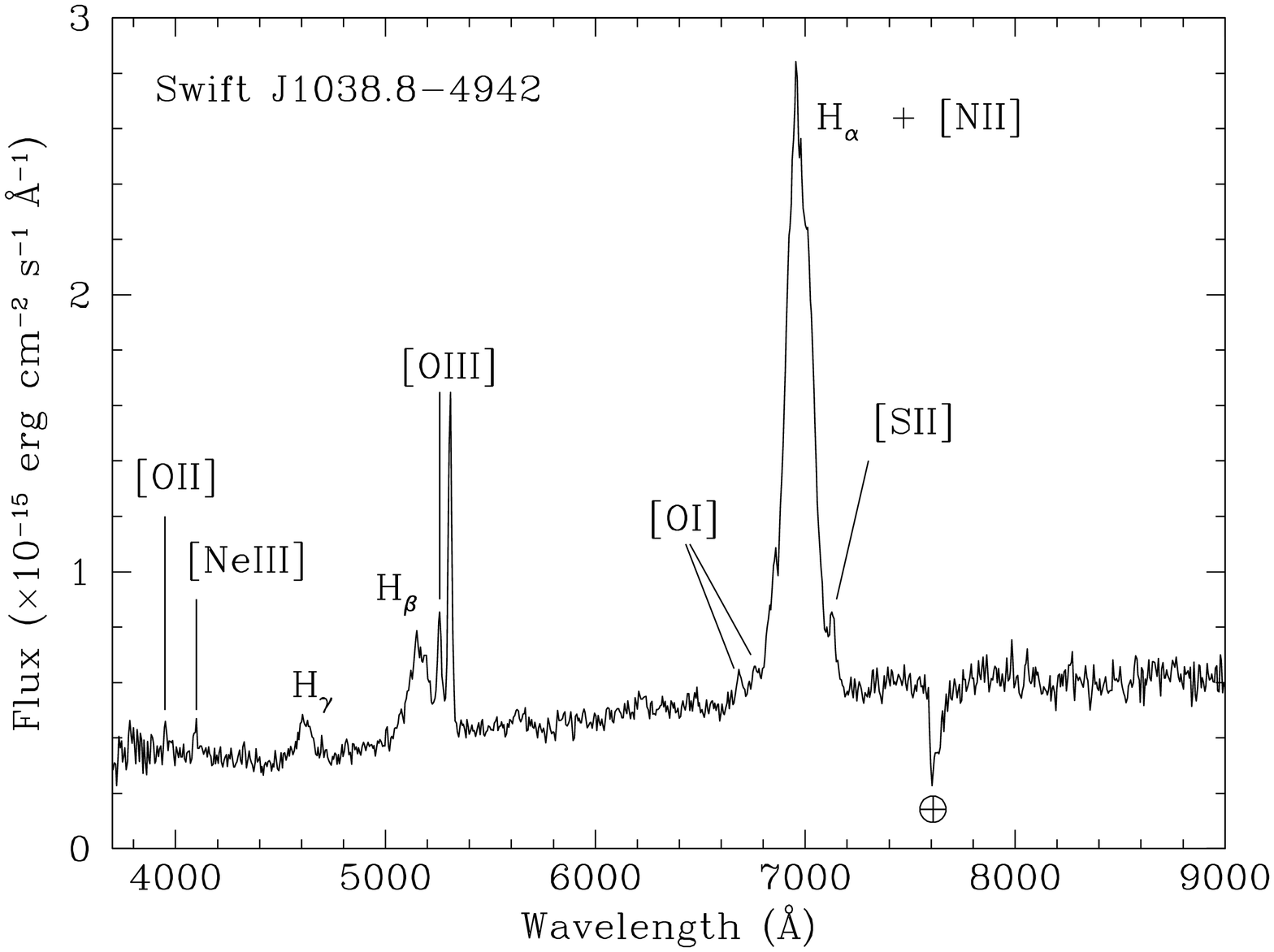}}

\caption{{\footnotesize Spectra (not corrected for the intervening Galactic absorption) of 
the optical counterparts of Swift J0444.1+2813 (upper left panel), Swift
J0601.9--8336 (upper right panel), Swift J0823.4--0457 (central left panel), Swift J0918.5+1618 
(central right panel), Swift J1009.3--4250 (lower left panel) and Swift J1038.8-4942 
(lower right panel). For each 
spectrum the main spectral features are labeled. The symbol $\oplus$ 
indicates atmospheric and telluric features.}}
\label{optical}
\end{figure}

\begin{figure}
\rotatebox{-90}{\epsscale{0.4}\plotone{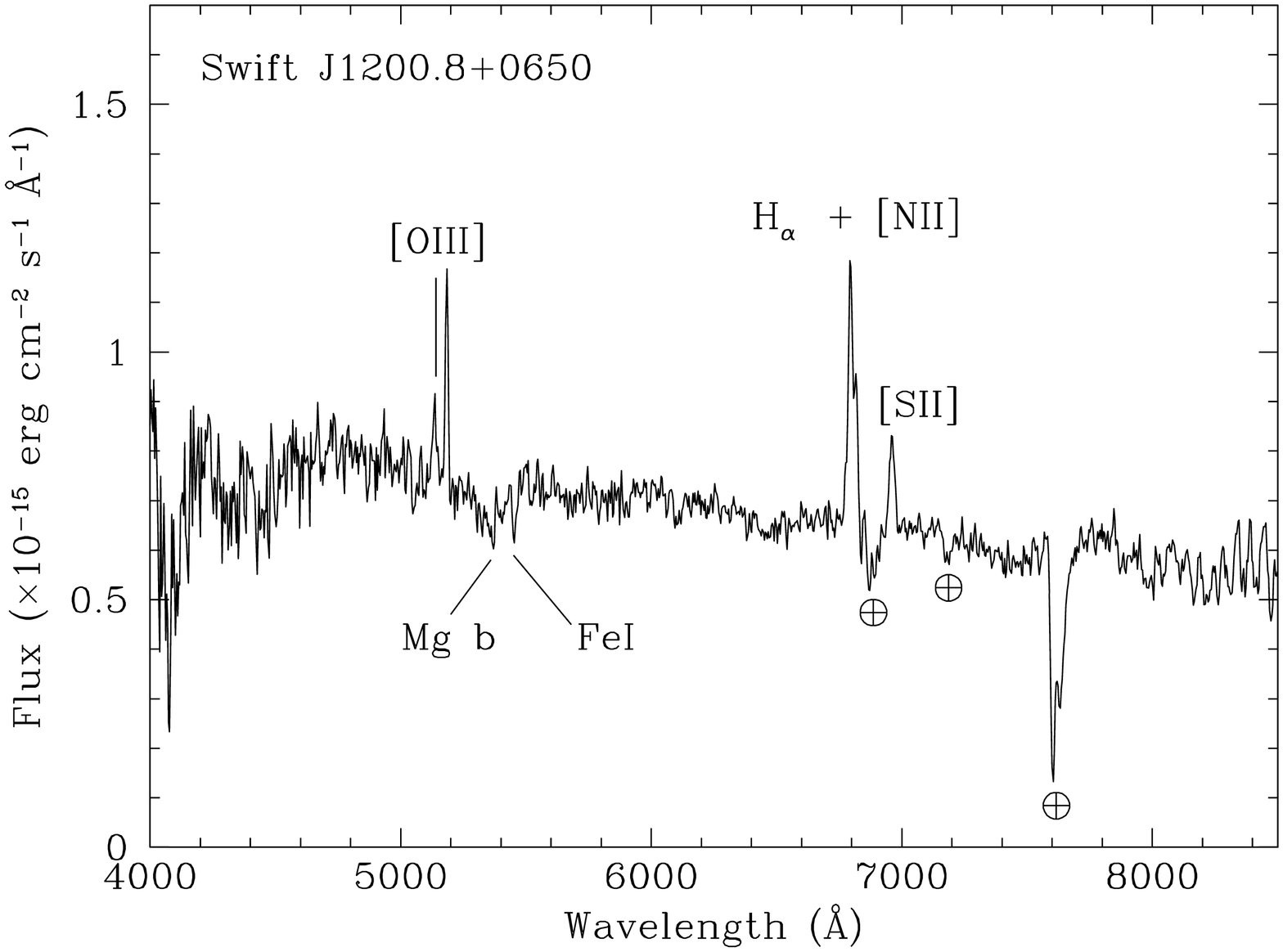}}
\rotatebox{-90}{\epsscale{0.4}\plotone{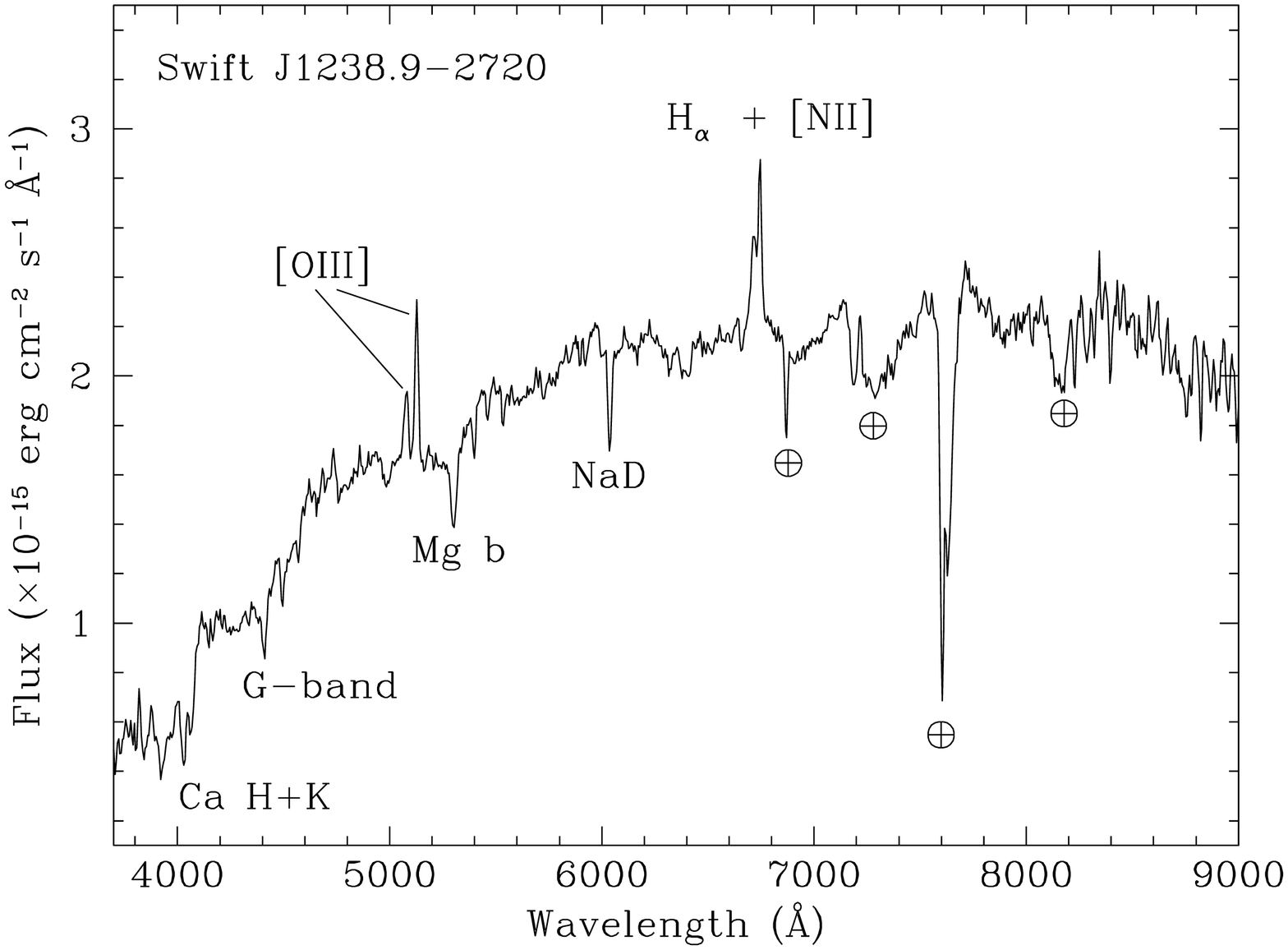}}

\rotatebox{-90}{\epsscale{0.4}\plotone{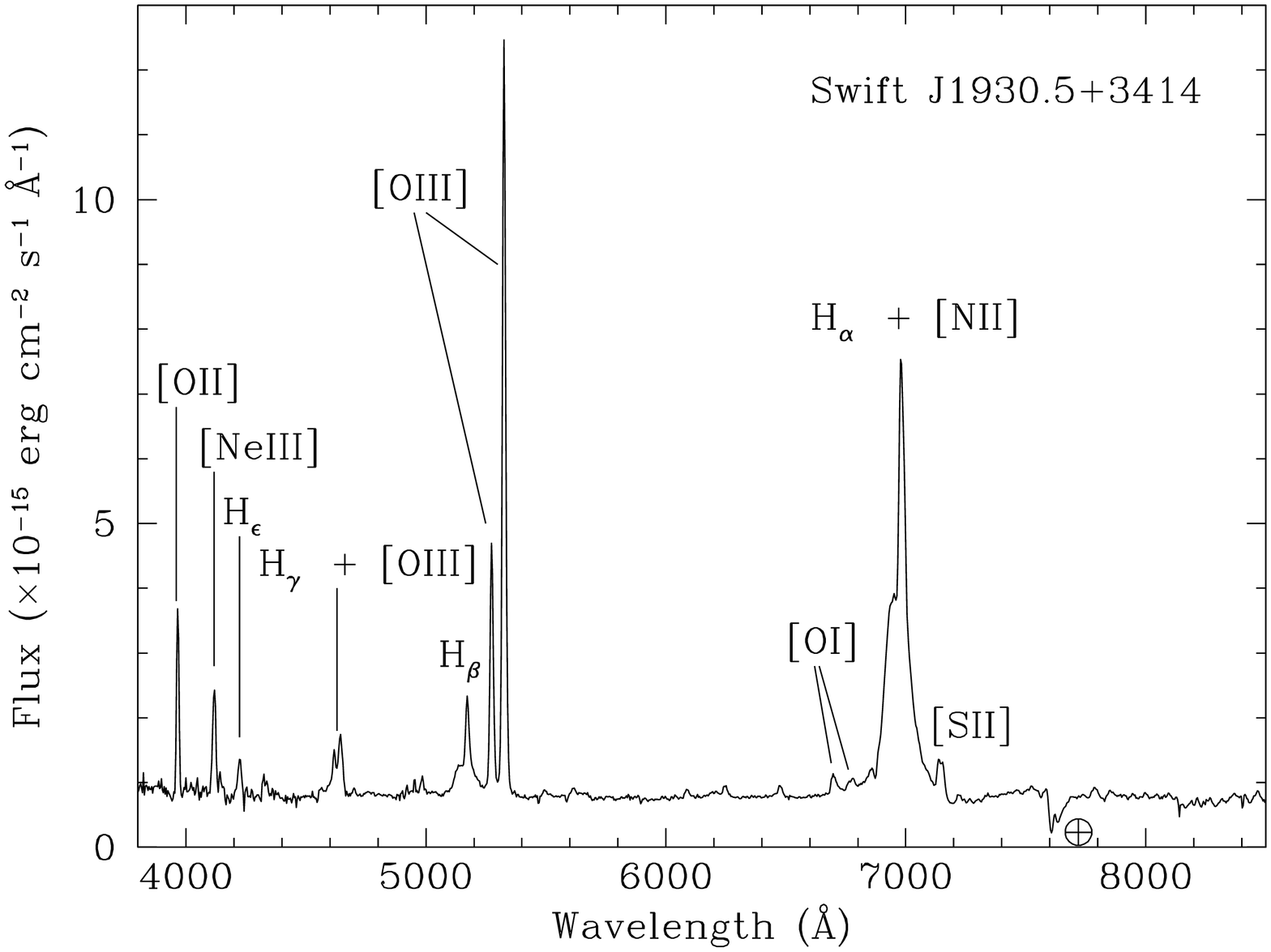}}
\rotatebox{-90}{\epsscale{0.4}\plotone{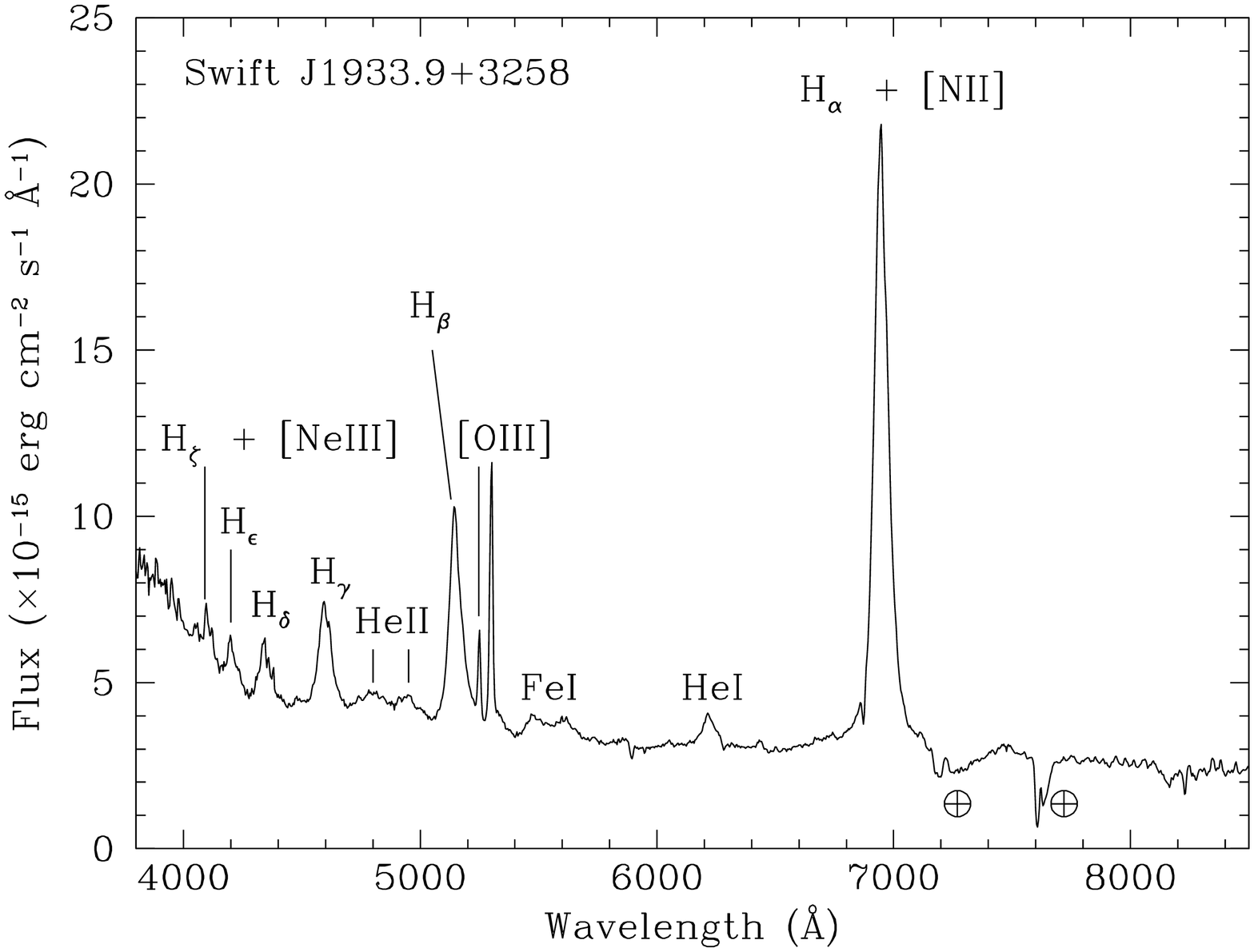}}

\rotatebox{-90}{\epsscale{0.4}\plotone{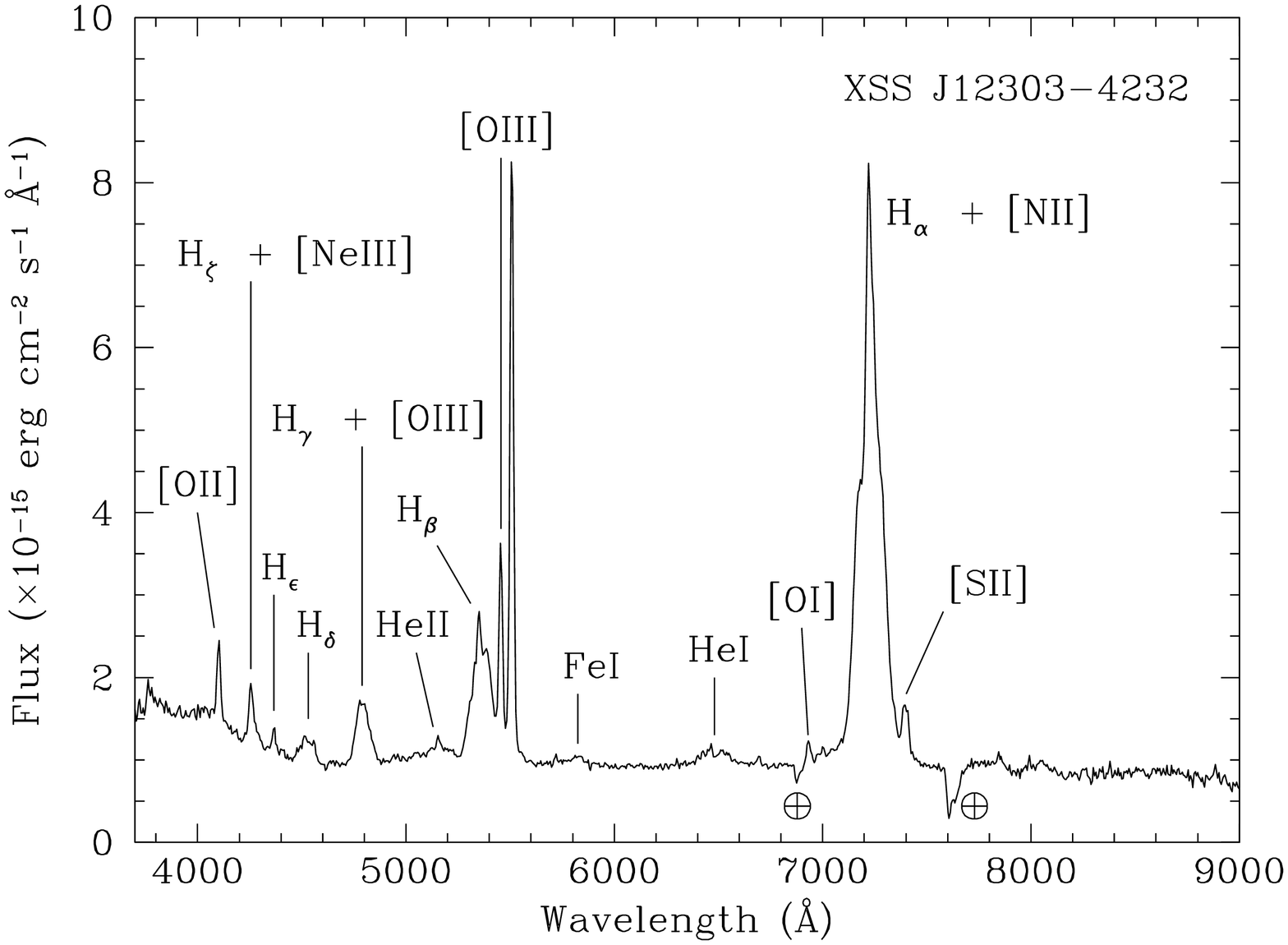}}

\caption{As Figure~\ref{optical}, but for objects 
Swift J1200.8+0650 (upper left panel),  
Swift J1238.9--2720 (upper right panel),
Swift J1930.5+3414 (central left panel), Swift J1933.9+3258
(central right panel) and XSS J12303--4232 (lower left panel).} 
\label{optical1}
\end{figure}

\begin{figure}
\rotatebox{-90}{\epsscale{0.7} \plotone{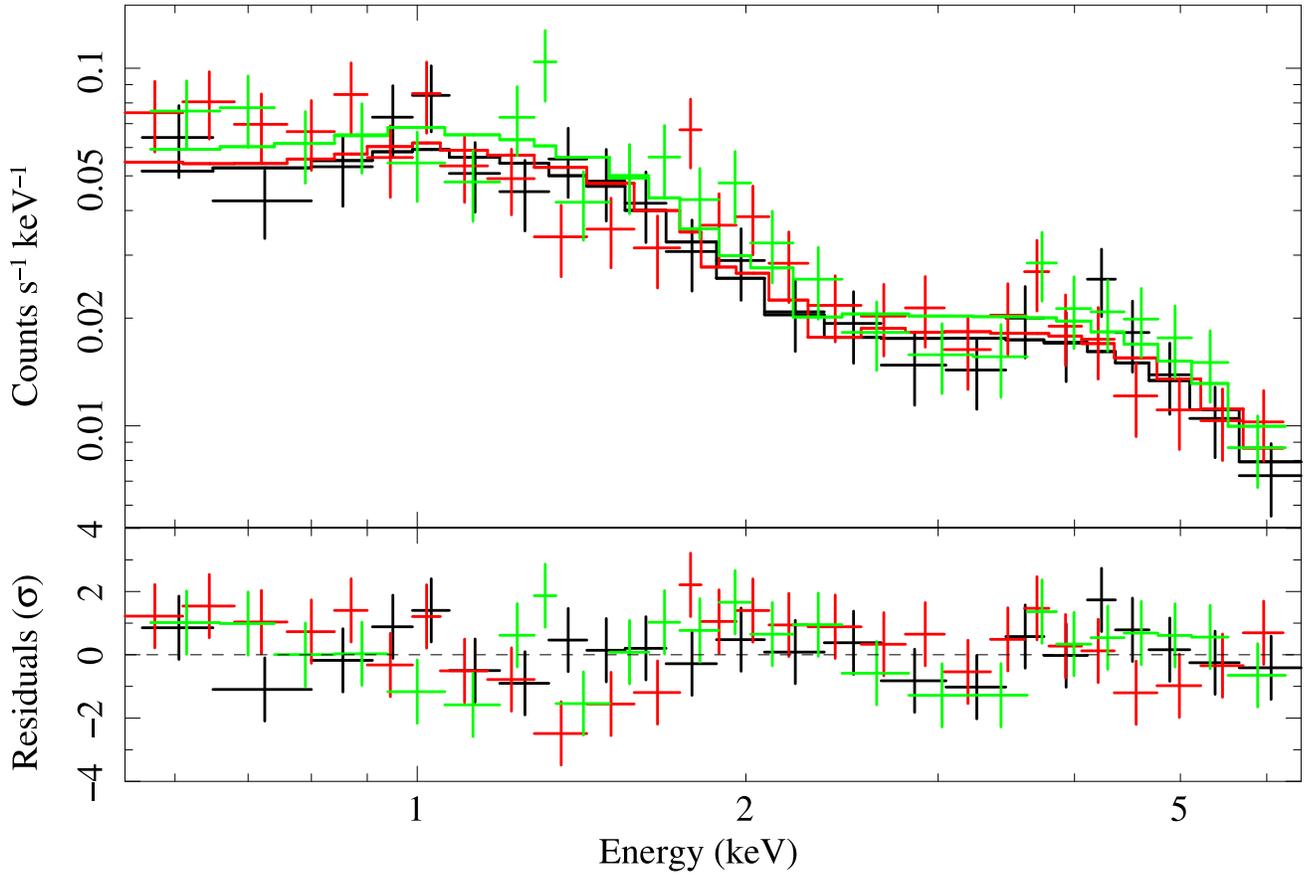}}
\caption{Three XRT spectra of Swift J0732.9--1331 fitted with a thermal component ({\sc mekal})
partially absorbed (upper panel); residuals to this model in units of $\sigma$ (lower panel).}
\label{spec0732}
\end{figure}

\begin{figure}
\rotatebox{-90}{\epsscale{0.7} \plotone{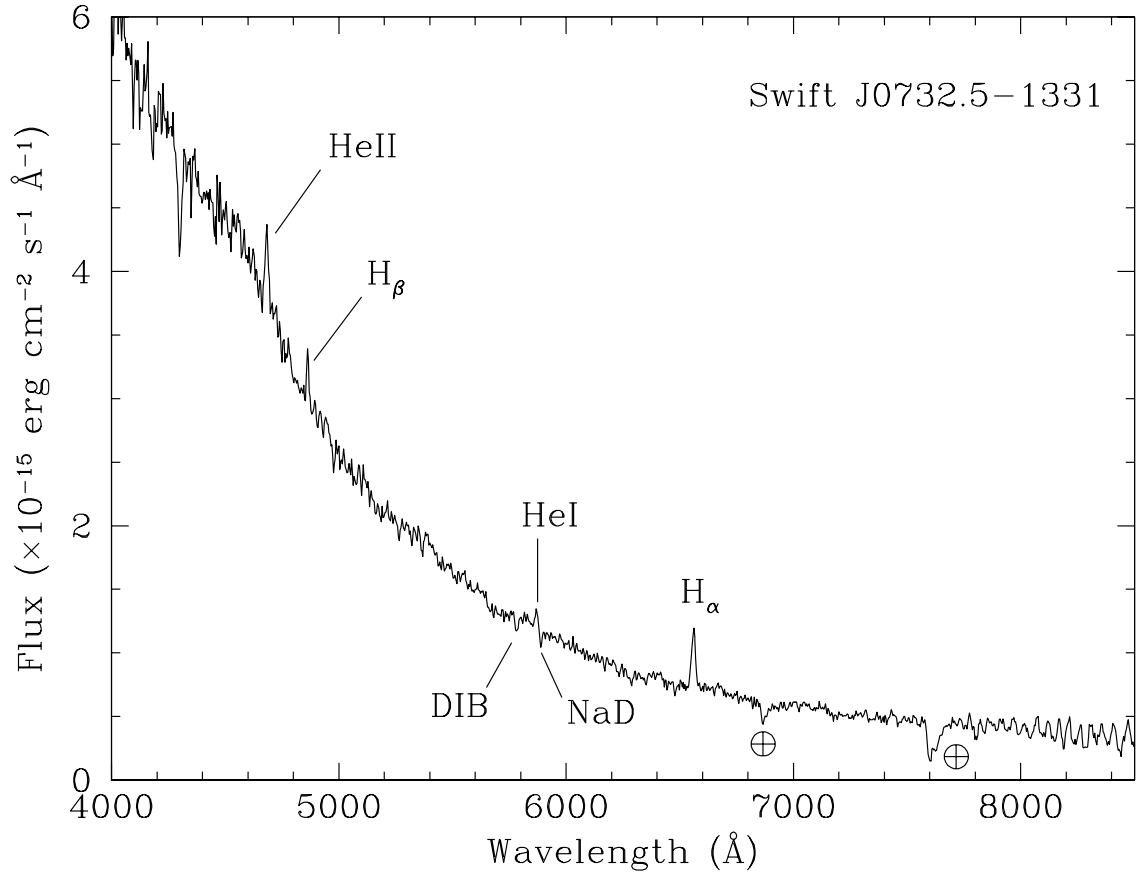}}
\caption{Spectrum (not corrected for the intervening Galactic absorption) of the optical  
counterpart of Swift J0732.9--1331. In the spectrum the main spectral features are labeled.
The symbol $\oplus$ indicates atmospheric and telluric features.}
\label{opt0732}
\end{figure}

Indeed, the optical spectrum we obtained (Fig.~\ref{optical}, central left panel)
shows that the $EWs$ 
of He {\sc ii} ($3.0\pm0.3$) \AA~and H$_\beta$ ($1.5\pm0.2$) \AA~have a 
ratio typical of IP CVs. In addition, the observed inverted 
H$_\alpha$/H$_\beta$ Balmer ratio ($\sim$2.2) is often found in this class 
of objects; it also suggests negligible absorption towards the source, in 
agreement with the X-ray data.

Considering no interstellar absorption, an absolute optical magnitude M$_V$ 
$\sim$9 and an intrinsic color index $(V-R)_0$ $\sim$0 mag (Warner 1995) 
for the object, and assuming that the true counterpart contributes to 
roughly one third of the total $R$ magnitude ($R$ $\sim$14.2) from the 
close star pair, we derive a distance of $\sim$190 pc. This implies a 
2--10 keV luminosity for this source of $\sim$$3.6\times10^{31}$ erg 
s$^{-1}$.

\clearpage

\begin{deluxetable}{lccccccc}
\tablecolumns{8}
\tabletypesize{\tiny}
\tablecaption{Log of the \emph{Swift} XRT observations presented in this paper.} 
\startdata
\hline
\hline
Source & R.A. &  Dec & Error & Obs date & Exposure & Energy band &Count rate \\
    &(J2000)  &(J2000)  & (arcsec) &  & (s)& (keV) & ($10^{-3}$ counts s$^{-1}$) \\
\hline
\hline
Swift J0444.1+2813 & 04 44 09.23 & +28 12 58.62 & 3.53 &  Aug 02, 2005 & 7206 & --  & --\\
Swift J0444.1+2813 &  &  &   & Aug 04, 2005  & 4759 & -- & --  \\
Swift J0444.1+2813 &  &  &   & Aug 11, 2005 & 6637 & -- & --  \\
Swift J0444.1+2813 &  &  &   & Aug 20, 2005 & 12793& 1.0--8.5 & $74.5\pm2.6$   \\ 
Swift J0444.1+2813 &  &  &   & Dec 05, 2005  & 29205& 1.0--9.3& $96.3\pm1.8$  \\
Swift J0444.1+2813 &  &  &   & Dec 06, 2005  & 24567& 1.0--8.0 & $80.4\pm1.8$  \\
Swift J0444.1+2813 &  &  &   & Dec 07, 2005  & 24124& 1.0--8.2& $87.2\pm1.9$  \\
\hline
Swift J0601.9--8636 & 06 05 39.24 & --86 37 51.32 & 4.57 & Dec 14, 2005 & 10211  
& 0.5--5.0 & $1.80\pm0.50$ \\
Swift J0601.9--8636 &   &  & &  Dec 18, 2005 & 2477  & -- & --  \\
Swift J0601.9--8636 &   &  & & Dec 20, 2005 & 700    & --  & --  \\
\hline
Swift J0732.5--1331 & 07 32 37.70 & --13 31 06.18&4.74& Jan 06, 2006 & 3445 & 0.5--6.4&$125.3\pm6.2$ \\
Swift J0732.5--1331 &  &  & & Jan 06, 2006 &332 & -- & --   \\
Swift J0732.5--1331 &  &  & & Apr 27, 2006&4044& 0.5--6.4 & $136.6\pm5.8$    \\
Swift J0732.5--1331 &  &  & & Oct 03, 2006 &4763 &0.5--6.4 & $146.3\pm6.8$  \\
\hline
Swift J0823.4--0457 & 08 23 00.93 & --04 56 04.01 
& 4.17 & Jan 06, 2006   & 1239 & 2.0--6.0 & $9.7\pm2.5$  \\
Swift J0823.4--0457 &   &  & & Oct 16, 2006 & 5362 & 2.0--6.0 & $13.1\pm1.5$   \\
Swift J0823.4--0457 &   &  & & Oct 21, 2006 & 4763& 2.0--6.0& $15.3\pm1.8$   \\
\hline
Swift J0918.5+1618  & 09 18 25.95 & +16 18 20.05
& 3.56 & Jan 06, 2006 & 673  & -- & -- \\
Swift J0918.5+1618  &   &  & & Jun 14, 2006 & 2258 & 0.7--6.4 & $352.6\pm12.6$ \\
Swift J0918.5+1618  &   &  & & Sep 28, 2006 & 5602 & 0.7--6.5 & $212.1\pm6.5$  \\
Swift J0918.5+1618  &   &  & & Jan 21, 2007 & 1750 & 0.7--7.0 & $294.5\pm13.1$  \\
\hline
Swift J1009.3--4250 & 10 09 48.12 & --42 48 42.60 
& 4.00& Jun 15, 2006 & 1809 & 0.5--6.5& $11.5\pm2.7$  \\
Swift J1009.3--4250 &   &  & & Jul 11, 2006 & 3877  & 0.5--7.5 & $13.8\pm2.0$  \\
Swift J1009.3--4250 &   &  & & Jul 13, 2006 & 1964  & 0.5--7.5 & $11.5\pm2.5$  \\
Swift J1009.3--4250 &   &  & & Jul 28, 2006 & 6180  & 0.2--7.5 & $18.4\pm1.8$  \\
Swift J1009.3--4250 &   &  & & Jul 30, 2006 & 0.91  & --& -- \\
\hline
Swift J1038.8--4942 & 10 38 44.87 & --49 46 52.73 & 3.56 & Oct 26, 2005    
& 5225 & 1.0--7.5 & $114.7\pm4.7$ \\
Swift J1038.8--4942 & & & & Nov 05, 2005 &  3852 & 1.0--7.0 & $90.2\pm4.8$  \\
Swift J1038.8--4942 & & &   & Dec 22, 2005 & 17112 & 1.0--8.0 & $81.0\pm2.2$  \\
\hline
Swift J1200.8+0650  & 12 00 57.74 & 06 48 21.04 & 3.64& Dec 11, 2005    &  15055
& 2.0--7.5 & $36.3\pm1.6$  \\
Swift J1200.8+0650  & & &   & Dec 21, 2005 & 3287 & 1.5--6.0 & $37.6\pm3.5$  \\
\hline
Swift J1238.9--2720 & 12 38 54.58 & --27 18 27.45 & 3.83  & Jun 15, 2005   &  8610
& 1.5--7.5 & $18.9\pm1.6$  \\
Swift J1238.9--2720 & & & &   Aug 15, 2005  & 2385 & 2.0--7.5 & $20.4\pm3.2$  \\
Swift J1238.9--2720 & & & & Aug 28, 2005 & 11527 & 2.0--8.0 & $17.0\pm1.3$  \\
\hline
Swift J1930.5+3414 & 19 30 13.70 & 34 10 52.74 & 4.15& Dec 12, 2005 & 3792 & -- & -- \\
Swift J1930.5+3414 &  & & & Dec 15, 2005 & 7363 & 1.0--9.0 & $11.8\pm1.4$   \\
Swift J1930.5+3414 &  &  &  & Dec 21, 2005 & 1312 & -- & --   \\
\hline
Swift J1933.9+3258 & 19 33 47.15 & 32 54 25.33 & 3.54 & Jul 07, 2006  & 1552
& 0.5--6.5 & $470.8\pm19.5$  \\
Swift J1933.9+3258 & & &  & Oct 13, 2006 & 5861  & 0.5--7.0 & $424.7\pm8.6$  \\
Swift J1933.9+3258 & & &  & Oct 15, 2006 & 4372  & 0.5--7.0 & $446.4\pm10.2$  \\
\hline
XSS J12303--4232   & 12 32 12.09 & --42 17 50.37 & 3.53 & Sep 08, 2005 & 15171
& 0.2--7.5 & $155.6\pm3.5$  \\
XSS J12303--4232   & & &  & Sep 12, 2005 & 16333 & 0.2--7.5 & $111.8\pm2.7$\\
\hline
\label{Tab1}
\enddata
\end{deluxetable}

\clearpage

\begin{deluxetable}{lcccccc}
\tablecolumns{7}
\tabletypesize{\tiny}
\tablecaption{XRT spectral analysis results of sources fitted with an absorbed 
power law. Frozen parameters are written between squared brackets.}
\startdata
\hline
\hline
\small
Source  &  $N_{\rm H_{\rm Gal}}^{a}$ & $N_{\rm H}^{a}$ &  $\Gamma$ &
Flux$^{b}$& $\chi^{2}/\nu$& $N_{\rm H}^{a,c}$\\
  &  &  &  &   (2--10 keV) &  & \\
\hline
\hline
Swift J0601.9--8636 (\# 1)&  0.111  & -- & [1.8] & $0.05\pm0.01$& 3.8/3 & -- \\
\hline
Swift J0823.4--0457 (\# 1)  & 0.0515 & -- & [1.8] & $4.2\pm1.1$& 0.1/2&25.2$^{+49.1}_{-25.2}$ \\
Swift J0823.4--0457 (\# 2)  & 0.0515 & -- & [1.8] & $5.3\pm0.6$& 4.1/4 &24.6$^{+13.6}_{-10.7}$ \\
Swift J0823.4--0457 (\# 3) & 0.0515 & -- & [1.8] & $5.8\pm0.7$& 2.4/4&33.1$^{+14.9}_{-11.2}$ \\
Swift J0823.4--0457 (all) &0.0515 & -- & [1.8] & $4.6\pm0.7$ & 7.8/13 &29.7$^{+9.0}_{-7.5}$  \\
\hline
Swift J0918.5+1618 (\# 2) & 0.0344 & $<$ 0.3 &1.80$^{+0.11}_{-0.08}$
&  $13.3\pm0.4$ & 31.4/34 & --\\
Swift J0918.5+1618 (\# 3)  & 0.0344 & 0.21$^{+0.12}_{-0.10}$ &
1.87$^{+0.20}_{-0.18}$ &  $8.8\pm0.3$ & 38.0/39 & 0.18$^{+0.06}_{-0.08}$\\
Swift J0918.5+1618 (\# 4)  & 0.0344 & $<$0.12& 1.92$^{+0.21}_{-0.19}$&
$10.0\pm0.4$ & 22.2/26 & $<$0.1\\
Swift J0918.5+1618 (2+3+4) & 0.0344 & 0.13$^{+0.06}_{-0.07}$ & 1.89$^{+0.12}_{-0.13}$ 
& $13.8\pm0.8$  & 98.2/97 & 0.10$^{+0.03}_{-0.05}$\\
\hline
Swift J1038.8--4942 (\# 1) & 0.272& 0.89$^{+0.38}_{-0.33}$ & 
1.25$^{+0.30}_{-0.29}$& $11.7\pm0.5$  & 26.7/27 & 1.50$^{+0.24}_{-0.20}$\\
Swift J1038.8--4942 (\# 2) & 0.272& 0.59$^{+0.47}_{-0.38}$ & 
1.41$^{+0.42}_{-0.38}$& $8.4\pm0.4$ & 8.3/14  & 1.20$^{+0.25}_{-0.22}$\\
Swift J1038.8--4942 (\# 3)& 0.272& 0.50$^{+0.17}_{-0.16}$ & 
1.16$^{+0.16}_{-0.10}$& $7.8\pm0.2$ & 29.8/40 & 1.10$^{+0.13}_{-0.14}$\\
Swift J1038.8--4942 (all)  & 0.272 & 0.50$^{+0.14}_{-0.13}$ & 
1.21$^{+0.15}_{-0.14}$ & $9.3\pm0.4$ & 91.0/106 & 1.28$^{+0.12}_{-0.10}$\\
\hline
Swift J1200.8+0650 (\# 1) &0.0144 & 6.50$^{+2.50}_{-1.92}$ & 1.26$^{+0.62}_{-0.53}$ &
$5.7\pm0.2$ & 27.7/24  & 8.08$^{+0.23}_{-0.98}$\\
Swift J1200.8+0650 (\# 2) &0.0144 & 7.94$^{+8.53}_{-4.62}$ & 1.87$^{+2.58}_{-1.67}$ &
$5.4\pm0.5$ & 11.8/11 & 7.65$^{+2.35}_{-1.64}$\\
Swift J1200.8+0650 (all) &0.0144 & 6.61$^{+2.17}_{-1.78}$ & 1.31$^{+0.58}_{-0.51}$ &
$5.8\pm0.3$  & 40.0/37 &8.30$^{+1.10}_{-0.90}$ \\
\hline
Swift J1933.9+3258 (\# 1) & 0.226 & $<$0.04 & 2.05$^{+0.14}_{-0.12}$& $14.4\pm0.6$
& 35.5/36& $<$ 0.02 \\
Swift J1933.9+3258 (\# 2) & 0.226 & $<$0.01 & 2.05$^{+0.07}_{-0.06}$& $13.5\pm0.3$
& 75.8/81& $<$ 0.01 \\
Swift J1933.9+3258 (\# 3) & 0.226 & $<$0.01 & 2.15$^{+0.08}_{-0.06}$& $12.6\pm0.3$
& 77.3/68 & $<$ 0.01 \\
Swift J1933.9+3258 (all)  & 0.226 & $<$0.04 & 2.08$^{+0.05}_{-0.04}$ & $13.8\pm0.4$ 
& 192.8/186 & $<$ 0.02 \\
\hline
\enddata
\label{Tab2}
\tablecomments{${^a}$ In units of $10^{22}$ cm$ ^{-2}$;\\
$^{b}$ In units of $10^{-12}$ erg cm$^{-2}$ s$^{-1}$;\\
$^{c}$ Assuming a photon index $\Gamma=1.8$;\\.}
\end{deluxetable}

\clearpage

\begin{deluxetable}{lcccccccccc}
\rotate
\tablecolumns{10}\tablecolumns{11}
\tabletypesize{\tiny}
\tablecaption{XRT spectral analysis results of sources requiring more complex models.
Frozen parameters are written between squared brackets.}
\startdata
\hline
\hline
\small
Source &   $N_{\rm H_{\rm Gal}}^{a}$ & $N_{\rm H}^{a}$ & $C_{\rm f}$& $\Gamma$ & $E_{\rm line}$&
$EW$ & $kT$ & Flux$^{b}$& $\chi^{2}/\nu$ & $N_{\rm H}^{a,c}$\\
  & &  &  & (keV)  & (eV)& (keV) &  (2--10 keV) & & & \\
\hline
\hline
Swift J0444.1+2813 (\# 4)  &0.196 &4.76$^{+1.30}_{-1.42}$ & --&  1.32$^{+0.35}_{-0.36}$$^{d}$ 
& -- & --  & -- &  $10.0\pm0.4$ & 48.5/49 & 6.37$^{+0.93}_{-0.74}$\\
Swift J0444.1+2813 (\# 5)  & 0.196&4.08$^{+0.62}_{-0.44}$ & -- &1.53$^{+0.20}_{-0.14}$$^{d}$ &
--  & -- &  -- & $12.0\pm0.2$ & 95.1/119 & 4.81$^{+0.36}_{-0.26}$\\
Swift J0444.1+2813 (\# 6)  &0.196 &4.31$^{+0.82}_{-0.74}$ & -- &1.42$^{+0.22}_{-0.19}$$^{d}$ &-- &  
-- & --  & $10.0\pm0.3$ & 97.6/113 & 5.40$^{+0.50}_{-0.40}$\\
Swift J0444.1+2813 (\# 7) &0.196 &4.21$^{+0.90}_{-0.83}$ & -- &1.41$^{+0.24}_{-0.19}$$^{d}$ &--&  
-- & -- & $11.0\pm0.3$ & 89.1/89 & 5.49$^{+0.50}_{-0.45}$\\
Swift J0444.1+2813 (all)  & 0.196 &4.25$^{+0.42}_{-0.32}$ &-- & 1.46$^{+0.11}_{-0.09}$$^{d}$ & --&-- 
& --  &  $10.7\pm0.3$ & 341.4/376 & 5.35$^{+0.65}_{-0.23}$\\
\hline
Swift J0732.5--1331 (\# 1)  & -- & 13.0$^{+12.7}_{-6.4}$& 0.68$^{+0.15}_{-0.32}$ & -- &  -- &  -- 
& 13.0$^{+66.0}_{-7.5}$& $10.2\pm0.4$ & 12.4/21 & --\\
Swift J0732.5--1331 (\# 3)  & -- & 5.1$^{+5.4}_{-2.2}$ & 0.61$^{+0.14}_{-0.23}$&-- &  -- &  -- 
& 9.8$^{+64.7}_{-4.3}$& $9.6\pm0.4$ & 32.2/24 & --\\
Swift J0732.5--1331 (\# 4)  & -- & 16.0$^{+16.2}_{-8.2}$&0.65$^{+0.15}_{-0.26}$&  -- &  -- &  --
& 21.9$^{+57.0}_{-13.9}$& $12.1\pm0.5$ & 24.4/21 & --\\
Swift J0732.5--1331 (all)  & -- &11.6$^{+6.3}_{-4.5}$ &0.63$^{+0.10}_{-0.19}$&  -- &  -- &  --  & 
15.4$^{+56.5}_{-5.91}$& $12.0\pm0.4$ & 73.7.72 & --\\
\hline
Swift J1009.3--4250 (\# 1) & 0.109& -- &-- & [1.8]$^{d}$ &-- & -- & -- & $1.9\pm0.4$ & 2.8/4 
&21.7$^{+195.5}_{-15.5}$ \\
Swift J1009.3--4250 (\# 2) & 0.109& -- &--& [1.8]$^{d}$ & -- & -- & --&$1.5\pm0.2$ & 12.5/16 
&14.3$^{+10.4}_{-5.6}$\\
Swift J1009.3--4250 (\# 3) & 0.109 & -- & --& [1.8]$^{d}$ & -- & -- & --&$1.9\pm0.4$ & 3.8/5 & 
45.7$^{+55.6}_{-21.5}$\\
Swift J1009.3--4250 (\# 4) & 0.109 &45.8$^{+9.5}_{-10.7}$ & --& 2.50$^{+0.70}_{-0.40}$$^{d}$ & 
-- & -- & -- &$2.8\pm0.3$ & 
23.0/20 &  41.4$^{+12.1}_{-8.8}$\\
Swift J1009.3--4250 (all) & 0.109 & 39.8$^{+10.8}_{-9.6}$ &--&  2.32$^{+0.58}_{-0.62}$$^{d}$ & 
-- & -- & -- & $2.0\pm0.3$ & 48.7/44 &  36.2$^{+8.6}_{-7.2}$\\
\hline
Swift J1238.9--2720 (\# 1) & 0.0668 & --&--&  [1.8] &
 6.48$^{+0.10}_{-0.13}$ & 528$^{+313}_{-295}$ & --& $4.4\pm0.4$ & 18.6/13 & 41.2$^{+10.5}_{-8.0}$\\
Swift J1238.9--2720 (\# 2) & 0.0668 & -- &--&  [1.8]  & 6.46$^{+0.20}_{-0.10}$  &
632$^{+600}_{-546}$  & -- & $5.4\pm0.8$  & 6.8/9 & 81.0$^{+51.3}_{-14.3}$\\
Swift J1238.9--2720 (\# 3)  & 0.0668 & -- &--&  [1.8] & 6.36$^{+0.10}_{-0.07}$ & 
432$^{+269}_{-232}$ & -- &  $4.2\pm0.3$ & 14.6/19 & 48.5$^{+15.2}_{-16.5}$ \\
Swift J1238.9--2720 (all)  & 0.0668 &--  &--&  [1.8] &
6.40$^{+0.07}_{-0.06}$  &508$^{+174}_{-200}$  & -- & $4.7\pm0.5$  & 46.5/42 & 46.3$^{+9.6}_{-7.8}$\\
\hline
Swift J1930.5+3414 (\# 2) & 0.172 & -- & --& [1.8]$^{d}$&-- & -- & -- & $1.8\pm0.2$ & 
10.4/17&28.5$^{+18.3}_{-11.2}$\\
\hline
XSS J12303--4232 (\# 1)  & 0.0726 & -- &--&  1.69$^{+0.06}_{-0.04}$ &
6.38$^{+0.08}_{-0.14}$  &331$^{+219}_{-203}$  & -- &  $5.4\pm0.1$& 98.8/109 & -- \\
XSS J12303--4232 (\# 2)  & 0.0726 & -- & --& 1.80$^{+0.06}_{-0.04}$ &
6.51$^{+0.07}_{-0.10}$   &360$^{+210}_{-203}$  & -- & $5.6\pm0.1$& 99.8/108& -- \\
XSS J12303--4232 (all) &  0.0726 & -- & --& 1.76$^{+0.03}_{-0.04}$ & 
6.42$^{+0.07}_{-0.04}$  &343$^{+115}_{-176}$  & -- & $5.4\pm0.1$& 208.0/219 & --\\
\hline
\enddata
\label{Tab3}
\tablecomments{${^a}$ In units of $10^{22}$ cm$^{-2}$;\\
$^{b}$ In units of $10^{-12}$ erg cm$^{-2}$ s$^{-1}$;\\
$^{c}$ Assuming a photon index $\Gamma=1.8$;\\
$^{d}$ In this case the best-fit model requires a second power law component, having the same photon 
index of the primary absorbed power law, and passing only through the Galactic column density.}
\end{deluxetable}

\clearpage

\begin{deluxetable}{cccccccccc}
\tablecolumns{10}
\tabletypesize{\tiny}
\tablecaption{\emph{ASCA} spectral analysis results of Swift J0918.5+1618 = Mrk 704.}
\startdata
\hline
\hline
 Energy band & Count rate &  $N_{\rm H_{\rm Gal}}^{a}$ & $N_{\rm H}^{a}$ & 
$Cf$&  $\Gamma$ &  $E_{line}$& $EW$ & Flux$^{b}$& $\chi^{2}/\nu$ \\
   (keV)& ($10^{-3}$ counts s$^{-1}$)   &  &    &  & & (keV) & (eV)&  (2--10 keV)  & \\
\hline
\hline
 1.2--10.0 &$66.8\pm1.1$  & 0.0344 & $13.5^{+4.9}_{-3.4}$& $0.64^{+0.10}_{-0.05}$& 
$1.55^{+0.19}_{-0.16}$&
$6.39^{+0.15}_{-0.14}$ & $233^{+111}_{-112}$ & $5.6\pm0.1$ & 193.7/191\\
\hline
\enddata
\label{Tab4}
\tablecomments{${^a}$ In units of $10^{22}$ cm$ ^{-2}$;\\
$^{b}$ In units of $10^{-12}$ erg cm$^{-2}$ s$^{-1}$.}
\end{deluxetable}

\clearpage

\begin{deluxetable}{lcccccc}
\tablecolumns{7}
\tabletypesize{\scriptsize}
\tablecaption{Log of the optical spectroscopic observations presented in this paper.}
\startdata
\hline
\hline
Source & \multicolumn{1}{c}{Date} & Telescope &
Mid-exposure & Grism or & Slit & Exposure \\
 & & & time (UT) & grating & (arcsec) & time (s) \\
\hline
\hline
Swift J0444.1+2813  & 01 Feb 2006 & 1.5m Loiano & 18:07:00 & \#4    & 2.0 & 2$\times$900  \\
Swift J0601.9--8636 & 22 Mar 2006 & 1.5m CTIO   & 00:06:53 & \#13/I & 1.5 & 900           \\
Swift J0732.5--1331 & 10 Feb 2006 & 1.5m Loiano & 21:43:16 & \#4    & 2.0 & 2$\times$1800 \\
Swift J0823.4--0457 & 09 Feb 2006 & 1.5m Loiano & 22:16:12 & \#4    & 2.0 & 2$\times$1800 \\
Swift J0918.5+1618  & 07 Feb 2006 & 1.5m Loiano & 22:42:45 & \#4    & 2.0 & 2$\times$1200 \\
Swift J1009.3--4250 & 25 Mar 2004 & 4m AAT         & 11:25:22 & 580V   & 6.7 & 1200          \\
Swift J1009.3--4250 & 25 Mar 2004 & 4m AAT         & 12:29:07 & 425R   & 6.7 & 600           \\
Swift J1038.8--4942 & 22 Mar 2006 & 1.5m CTIO   & 02:24:02 & \#13/I & 1.5 & 2$\times$600  \\
Swift J1200.8+0650  & 10 Feb 2006 & 1.5m Loiano & 01:25:50 & \#4    & 2.0 & 2$\times$1800 \\
Swift J1238.9--2720 & 23 Mar 2006 & 1.5m CTIO   & 01:46:56 & \#13/I & 1.5 & 2$\times$1200 \\
Swift J1930.5+3414  & 05 Aug 2006 & 1.5m Loiano & 00:13:45 & \#4    & 2.0 & 2$\times$1800 \\
Swift J1933.9+3258  & 28 Jul 2006 & 1.5m Loiano & 01:08:14 & \#4    & 2.0 & 2$\times$900  \\
XSS J12303--4232    & 04 Apr 2006 & 1.5m CTIO   & 00:47:38 & \#13/I & 1.5 & 2$\times$900  \\
\hline
\enddata
\label{Tab5}
\end{deluxetable}

\clearpage

\begin{deluxetable}{lcccccccccc}
\rotate
\tablecolumns{11}
\tabletypesize{\tiny}
\tablecaption{Fluxes of 
H$_\alpha$, H$_\beta$, [O {\sc iii}] and [N {\sc ii}] emission lines detected in 
the spectra of the objects reported in Figures 14 and 15.
Measurements and upper limits are reported at 1$\sigma$ and 3$\sigma$ 
confidence levels, respectively.
For the extragalactic objects the values are corrected for Galactic 
reddening assuming a color excess, $E(B-V)_{\rm Gal}$, as per Schlegel et 
al. (1998) and which is reported in the Table as well as the optical ($A_{\rm V_{\rm opt}}$)
and X-ray ($A_{\rm V_{\rm X}}$) extinctions.}
\startdata
\hline
\hline
Source & $F_{\rm H_\alpha}$$^{a}$ & \multicolumn{2}{c}{$F_{\rm H_\beta}$$^{a}$} &
$F_{\rm [OIII]}$$^{a}$ & $F_{\rm [NII]}$$^{a}$& $E(B-V)_{\rm Gal}$ & 
 $A_{\rm V_{\rm opt}}$ & $A_{\rm V_{\rm X}}$$^{\dag}$ & Type\\
  &   & Broad  & Narrow & & & & & &  &  \\
\hline
\hline
Swift J0444.1+2813  & $7.4\pm0.7$   & --  & $<$0.1 & $4.4\pm0.7$ & $11.0\pm1.1$  & 0.856
& $>$10 & 19--24 & Seyfert 2$^{b}$ \\ 
Swift J0601.9--8636 & $1.57\pm0.18$ & --  & $<$0.041 & $<$0.5      & $1.52\pm0.18$ & 0.140
& $>$8.1 & $<$0.5 & Seyfert 2$^{b}$ \\
Swift J0732.5--1331 & $1.0\pm0.1$   &\multicolumn{2}{c}{$4.4\pm0.5^{d}$}  & --  & -- & -- 
& -- & -- & IP CV$^{b,c}$ \\
Swift J0823.4--0457 & $1.90\pm0.13$ & -- &$0.31\pm0.09$ & $2.0\pm0.1$ & $1.56\pm0.10$ &0.046
& 2.4 & 134 &Seyfert 2$^{b}$ \\
Swift J0918.5+1618  & *             & $15.4\pm0.8$ & $1.15\pm0.12$  & $7.2\pm0.5$  & * & 0.029  
&-- & $>$4.5 & Seyfert 1.2\\
Swift J1009.3--4250 & $2.4\pm0.1$   & -- & $0.60\pm0.05$ & $9.0\pm0.1$ & $3.30\pm0.13$ & 0.167
&1.1 & 163--179 & Seyfert 2 \\
Swift J1038.8--4942 & *             & \multicolumn{2}{c}{$18.3\pm1.3^{d}$} & $11.1\pm0.6$ & * & 0.496
& -- & 2--6 &Seyfert 1.5$^{b}$ \\
Swift J1200.8+0650  & $1.17\pm0.12$ & --  & $<$0.25  & $0.58\pm0.04$ & $0.42\pm0.06$ & 0.017
& $>$1.6 & 30--37 & Seyfert 2$^{b}$ \\
Swift J1238.9--2720 & $1.61\pm0.16$ & --  & $<$0.07  & $1.59\pm0.11$ & $1.43\pm0.14$ & 0.072
& $>$6.4 & 208 & Seyfert 2$^{b}$ \\
Swift J1930.5+3414  & *             & $8.5\pm1.3$   & $2.7\pm0.4$  & $33\pm1$ & *   & 0.187 
& -- & 128 & Seyfert 1.5--1.8\\
Swift J1933.9+3258  & *             & \multicolumn{2}{c}{$92\pm12^{d}$} & 26$\pm$2 & * & 0.271
& -- & $<$0.1 & Seyfert 1.2 \\
XSS J12303--4232    & *             & $20\pm2$ & $2.2\pm0.2$  & $20\pm1$ & *   & 0.104 
& -- & -- & Seyfert 1.5$^{b}$\\
\hline
\enddata
\label{Tab6}
\tablecomments{${^a}$ In units of $10^{-14}$ erg cm$^{-2}$ s$^{-1}$;\\
$^{b}$ This source has been classified in this work for the first time;\\
$^{c}$ IP CV (Intermediate Polar (magnetic) Cataclysmic Variable;\\
$^{d}$ Sum of the fluxes of broad and narrow components;\\
$^*$ In this case [N {\sc ii}] and H$_\alpha$ are heavily blended;\\
$^{\dag}$ This value has been obtained using the conversion $A_{\rm V_{\rm X}} \simeq
\frac{N_{\rm H}}{2.22\times10^{21}}$ (Zombeck 1990)}
\end{deluxetable}

\clearpage

\begin{deluxetable}{lcccccc}
\tablecolumns{7}
\tabletypesize{\scriptsize}
\tablecaption{Hard X-ray and infrared fluxes, and diagnostics luminosity ratios.}
\startdata
\hline
\hline
Source  &$F_{\rm HX}^{a,b}$  &$F_{\rm IR}^{b}$&  $L_{\rm X}$/$L_{\rm [OIII]}$ & 
$L_{\rm X}$/$L_{\rm IR}$ &  $L_{\rm [OIII]}$/$L_{\rm IR}$ & $L_{\rm X}$/$L_{\rm HX}$\\
  &  (10--100 keV)  &    &    &   &   &   \\
\hline
\hline
Swift J0444.1+2813   & 6.9$^{1}$ & 8.3  & $<$0.008 & 0.13  & $>$16.9 &  0.16\\
\hline
Swift J0601.9--8636  & 3.7$^{1}$ & 42.90 &    --    & 0.00012   & -- & 0.001 \\
\hline
Swift J0823.4--0457$^{c}$ & 2.7$^{2}$& $<$7.17   &   17.5   & $\le$0.06 & $>$0.0033  & 0.16\\
\hline
Swift J0918.5+1618   & 2.7$^{1}$ &8.16   &  109     & 0.096     & 0.0009 & 0.29 \\
\hline
Swift J1009.3--4250  & 2.5$^{3}$ & --    &    7.4   &  --       &  -- & 0.08  \\
\hline
Swift J1038.8--4942  & 3.4$^{1}$ & --    &   83.8   &  --       & -- & 0.27 \\
\hline
Swift J1200.8+0650   &  2.6$^{1}$&$<$3.6    & $<$200   & 0.16      & $>$0.0008 & 0.22\\
\hline
Swift J1238.9--2720  & 12.4$^{1}$ & 5.7    & $<$0.39  & 0.08      & $>$0.21 & 0.04\\
\hline
Swift J1930.5+3414   &  2.5$^{1}$&$<$6.5    &    5.5   & $>$0.028  & $>$0.0051  & 0.07 \\
\hline
Swift J1933.9+3258   &  2.7$^{4}$ & 6.4    &   30.4   & 0.12       & 0.0041 & 0.29 \\
\hline
XSS J12303--4232$^{\dag}$  &  0.61   & 4.2    &   24.0   & 0.11      &  0.0047 & 0.89 \\
\hline
\enddata
\label{Tab7}
\tablecomments{${^a}$ References: (1) BAT AGN catalog (available at 
{\tt http://www.astro.umd.edu/~lwinter/research/AGN.html});
(2) Aiello et al. 2006a; (3) Ajello et al. 2006b; (4) Grupe et al. 2006;\\
$^{b}$ In units of $10^{-11}$ erg cm$^{-2}$ s$^{-1}$;\\
$^{c}$ Assuming that IRAS 08205--0446 is associated with 
Fairall 272;\\
$^{\dag}$ For this source we consider the \emph{RXTE} 8--20 keV flux.}
\end{deluxetable}

\clearpage

\clearpage

\clearpage

\clearpage

\clearpage

\clearpage

\clearpage

\clearpage

\clearpage

\clearpage

\clearpage

\clearpage

\clearpage

\clearpage

\clearpage

\clearpage

\end{document}